\documentclass[draft,showpacs,preprintnumbers,amsmath,amssymb]{revtex4-1}
\newcommand{\fant}[1]{\phantom{#1}}
\newcommand{\be}{\begin{equation}}
\newcommand{\ee}{\end{equation}}
\newcommand{\wdg}{\wedge}
\newcommand{\ot}{\otimes}
\usepackage{multirow}

\begin{document}
\begin{abstract}
Quadratic curvature gravity  equations are projected to a complex null coframe by using the algebra of exterior forms and expressed in terms of the  spinor quantities  defined originally by Newman and Penrose. As an application, a new family of impulsive gravitational wave  solutions  propagating  in various Petrov type D  backgrounds is introduced.
\end{abstract}

\title{Gravitational wave solutions of quadratic curvature gravity using a null coframe formulation}

\author{Ahmet Baykal}
\email{abaykal@nigde.edu.tr\\failure2communicate@gmail.com}
\affiliation{Department of Physics, Faculty of Arts and Sciences, Ni\u gde University,  Bor Yolu,  51240 Ni\u gde, Turkey}

\date{\today}

\maketitle

\section{Introduction}

Although there is  convincing  indirect evidence for the existence of gravitational waves \cite{hulse-taylor}, they have not yet been detected directly. Currently, there are continuing efforts to directly detect gravitational radiation \cite{gw-observation}. The detection will open a new window for the observation of astrophysical phenomena on  cosmological scale. In parallel to these efforts, theoretical investigations of the properties of the gravitational radiations   are also of considerable interest in the framework of general theory of relativity  or  in the framework of any viable alternative, including a modification of the general relativity theory (GR).

A natural mathematical framework  for a theoretical investigation of radiative solutions, and in general algebraically special solutions of the Einstein field equations, is the spinor formulation of GR by introducing a complex null coframe by Newman and Penrose \cite{newman-penrose,exact-sol-stephani,griffiths-podolsky-exact-sol}. A similar null coframe formulation for a modified gravitational model may also have some theoretical value that may provide a convenient formulation in the theoretical search for radiative solutions and their properties in these theories. Such a null coframe formulation has the potential to facilitate  the  comparison of algebraically special  solutions to the modified  theories with those of GR.

According to the peeling-off theorem \cite{peel}, along a null direction, the leading term for the gravitational radiation in the radiation zone is of Petrov  type N,  and therefore its properties  are relevant in an attempt to observe the gravitational radiation from distant astrophysical sources. For example, distinct properties of type the N field for particular gravitational models may provide  viability criteria in light of  future observational data, e.g., the amplitude  correction resulting from quadratic curvature (QC) interactions, (see for example, \cite{neto-perturbative}).

Plane-fronted gravitational waves with parallel rays,  $pp$-wave metrics for short, are exact solutions of Einstein field equations found quite a long time ago \cite{brinkman,ehlers-kundt} and they have the peculiar property that they linearize the field equations. They are of Petrov type N and the corresponding Weyl tensor has fourfold null eigenvector \cite{chandra}. $pp$-wave metrics belong to the more general Kundt  family of metrics \cite{Kundt}, characterized by null geodesic congruence that has vanishing optical  scalars, i.e., a shear-, expansion-, and twist-free geodesic null congruence. As the defining property, $pp$-wave metrics require the null geodesic congruence  to be covariantly constant. However, there are twisting type N solutions \cite{hauser,chinea} that do  not belong to the Kundt  family of metrics, or there are type N metrics that have expanding null geodesic congruence which belong to the  Robinson-Trautman family of metrics.

There are several methods to construct the gravitational wave solutions introduced above. For example,
by boosting the Schwarzchild solution to the speed of light and at the same time reducing its mass to zero in an appropriate way
one obtains Aichelburg-Sexl solutions corresponding to a null particle. Later, the boost method is applied to the other black hole solutions as well. In another direction, for null particles  having  sources with multipole structure,  $pp$-wave metrics are constructed in
\cite{griffiths-podolsky}.
Later,  $pp$-wave solutions with a cosmological constant  were introduced \cite{hotta-tanaka,podolsky-griffiths} by extending the solutions of  Aichelburg and Sexl by boosting the Schwarzchild--de Sitter solution to the speed of light. The metrics with vanishing curvature invariants  are of the Kundt family with admissible  Petrov types of III, N or O as discussed in \cite{podolsky-ortaggio}. Previously, $pp$-wave metrics with various Petrov type D backgrounds  have been studied in \cite{ortaggio-podolsky,ortaggio}. The present work can be considered as an extension of the works \cite{ortaggio-podolsky,ortaggio} to the models based on the  gravitational actions involving  the general quadratic curvature terms in four dimensions.

Another method of obtaining $pp$-wave solutions is  Penrose's  geometrical construction of the cut and paste method  where Minkowski spacetime is cut along a null cone and then the two pieces are reattached with a warp. For further details of the methods for constructing impulsive waves, see \cite{griffiths-podolsky}.

$pp$-wave metrics also appear to be   exact  solutions to other important geometrical theories as well. They are solutions to the string theory in all orders of the string tension \cite{string}. Familiar $pp$-wave metrics also constitute a set of solutions common in Brans-Dicke and general relativity theories \cite{tupper}. In the context of metric-affine gravity, $pp$-wave solutions are generalized to spacetimes  equipped with a metric-compatible connection having nonzero torsion \cite{obukhov-metric-affine-pp-waves,pasic}. A complex null coframe formalism constructed from an orthonormal coframe, in the spirit of the present work, was used to study plane waves in supergravity theories \cite{beler-dereli}.

The exact solutions to Weyl conformal gravity, which is a particular type of QC  theory, are studied with the $(2,2)$ decomposability assumption of the metric in \cite{fiedler-schimming,hj-schmidt}. Using Newman-Penrose (NP) spin coefficient formalism, the geodesic deviation equation for the   $pp$-wave metrics has been studied in the context of QC gravity \cite{neto-geodesic-deviation}. Recently,  AdS-wave solutions to QC gravity have been given in \cite{ads-bayram-tekin} using a Kerr-Schild type metric Ansatz. At the same time, a family exact solutions to QC gravity belonging to types N and III according to the algebraic classification of Weyl tensor in higher dimension \cite{pravda-pravdova-coley-milson} is presented in \cite{malek-pravda}. This paper deals with QC gravity in four spacetime dimensions and impulsive wave solutions to QC gravity along the lines of  \cite{malek-pravda}.

The outline of the paper is as follows.  In the following section, the metric field equations that follow from the general quadratic curvature gravity in four spacetime dimensions are formulated relative to a rigid (i.e. an orthonormal  or a null) coframe in terms of some auxiliary tensor-valued forms.  The null coframe formulation is then  applied to a study of $pp-$wave  solutions in various algebraically special backgrounds in the ensuing sections.  In particular, in Sec. III, the solutions of various type D background spacetimes of the direct product form for the QC action are introduced.  Subsequently, using the same metric Ansatz and the coframe associated with the frame fields of \cite{ortaggio}, a new family of impulsive wave solutions  in direct product background spacetimes  for QC gravity is discussed and a general fourth-order partial differential  equations for the  profile function are derived. The impulsive wave solutions can be considered as extensions of the solutions (introduced in \cite{ortaggio}) to general QC gravity in parallel to the recent gravitational wave  solutions in \cite{malek-pravda}. The paper ends with some  general remarks regarding the extensions and applications of the null coframe formalism developed.

The essential technical details for the null coframe formalism  are given in the Appendix.  As a novel technical feature  in the exposition of the  NP field equations in terms of exterior forms, the Appendix includes a general recipe for calculation of  the complex NP spin coefficients and curvature spinors. The algebra of exterior forms used in the present work follows closely the formalism introduced previously in \cite{baykal} for general QC gravity formulated in terms of auxiliary tensor-valued forms. The notation for the null coframe version of the formalism  is established with the help of the formulas in the  Appendix and in particular,  the QC gravity equations are adapted  to a null coframe which are expressible either in terms of the original NP spinor quantities \cite{newman-penrose} or as tensorial expressions by making use of Einstein summation convention.

\section{QC field equations relative to a rigid coframe}

In this section the field equations for the QC models coupled to Maxwell field will be derived in a form  that is convenient to study the formulation of the metric field equations relative to  a rigid coframe, e.g., an NP null complex coframe or an orthonormal real coframe.

The metric equations in such a formulation can be derived from a coframe variational derivative of the action by using the first order formalism \cite{kopczynsky}.  In this formalism, one starts with independent  connection and the coframe 1-forms as independent gravitational variables and the metric theory is recovered by constraining  the connection 1-forms to be Levi-Civita by introducing  appropriate Lagrange multiplier terms  and imposing the  metric compatibility conditions. The general framework of presentation here closely follows the one given in \cite{kopczynsky,hehl-mccrea-mielke-neemann} and for further in formation on the notation for the exterior differential forms adopted, the reader is also referred to \cite{straumann,benn-tucker}.
In particular, the formulation of QC field equations in terms of the auxiliary tensor-valued forms defined below, relative to a local orthonormal coframe, is  directly taken from \cite{baykal} without a change except that the  signature of the metric here is taken  to be mostly minus in order to conform with the original definitions of NP quantities.

The geometrical definitions and conventions belonging to a null coframe in the following sections are described in the Appendix.
The null coframe formulation  of Einstein field equations using the exterior algebra of complex null coframe has been used
before, for example in \cite{debney-kerr-schild,guven-yoruk,bilge-gurses,exact-sol-stephani}. The formulation below extends the null coframe formulation of GR to the general QC gravity  allowing one to obtain spinor expressions directly from the corresponding tensorial expressions.
Such a formulation, however, requires the metric field equations to be formulated in terms of differential forms.

\subsection{Total variational derivative by constrained first-order formalism}

The current paper discusses the gravitational model based on the  Einstein-Maxwell model extended by general quadratic curvature terms and a cosmological constant. Maxwell field will be taken as the only matter field minimally coupling to the gravitational action. It is well known that the most general QC Lagrangian  density can be conveniently be expressed in terms of the differential forms   $R^2*1, R_\alpha\wdg*R^\alpha$.
The  gravitational theory coupled to the Maxwell field  studied below is described by the Lagrangian density
\begin{align}
\mathcal{L}
&=
\mathcal{L}_g[\theta^\alpha, \omega^{\alpha}_{\fant{a}\beta}, \Omega^{\alpha}_{\fant{a}\beta}]+\mathcal{L}_m[F, \theta^\alpha]
\nonumber\\
&=
\frac{1}{\kappa^2}\left(\tfrac{1}{2}R*1
+
\Lambda_0*1
\right)
+
\frac{a}{2}R^2*1+\frac{b}{2} R^\alpha\wdg *R_\alpha-\frac{1}{2}F\wdg *F.
\label{lag-def}
\end{align}
The total Lagrangian density $\mathcal{L}$ is proportional to volume four form  in four dimensional pseudo-Riemannian  manifold  where $\Lambda_0$ stands for the cosmological constant and  $a,b$ are constant coupling parameters for the QC terms. The  last term on the right-hand side is the Maxwell Lagrangian 4-form
$\mathcal{L}_m[F, \theta^\alpha]$, expressed in terms of the Faraday 2-form $F=\tfrac{1}{2}F_{\alpha\beta}\theta^{\alpha\beta}$.

The field equations then follow from the total variational derivative of the action integral
\be\label{action-def}
I=
\int_{U}\mathcal{L}
\ee
defined on a compact submanifold $U\subset M$ of a pseudo-Riemannian manifold $M$. The variational derivative of a geometrical object is denoted by a
$\delta$ and it  commutes with the exterior derivative and integration as well.

In order to derive the variational derivative $\delta I=0 $ that follows from the action (\ref{action-def}) in a suitable form, the first order formalism is used as a convenient mathematical framework. In this  formalism the coframe and connection 1-forms are taken as independent gravitational variables and the exterior derivatives of these variables only present in the gravitational sector therefore, the matter fields are assumed to couple minimally.
The formalism can be used to study a quite wide  range of gravitational subtheories introducing  appropriate constraints by adopting either orthonormal coframe \cite{kopczynsky} or a coordinate coframe \cite{safko-elston}. For the present work, the former is suitable and the gravitational field equations  for the pseudo-Riemannian metric can be derived from the coframe variation of the Lagrangian (\ref{lag-def}) subject to the constraints that the torsion tensor $\Theta^\alpha=D\theta^\alpha$ vanishes and the torsion free connection is the metric compatible. Relative to any rigid coframe, (e.g., an orthonormal coframe, null coframe or a half-null coframe) where the metric components are constants, the metric compatibility condition for the connection 1-form reads $D\eta_{\alpha\beta}=-\omega_{\alpha\beta}-\omega_{\beta\alpha}=0$. It is an advantage of adopting a  rigid coframe that, this algebraic condition for the connection components can be implemented into the variational derivative by anti-symmetrization of the coefficients of $\delta\omega_{\alpha\beta}$ term, corresponding to the antisymmetry of the derivative ${\partial \mathcal{L}}/{\partial\omega_{\alpha\beta}}$, whereas the torsion-free condition $\Theta^\alpha=0$   can be imposed by extending the original Lagrangian density to include  Lagrange multiplier 2-forms $\lambda^\alpha=\frac{1}{2}\lambda^\alpha_{\fant{a}\mu\nu}\theta^{\mu\nu}$ in the form
\be\label{extended-lag-def}
\mathcal{L}_e[\theta^\alpha, \Omega^{\alpha}_{\fant{a}\beta}, \Theta^\alpha, \lambda_\alpha, F]
=
\mathcal{L}
+
\mathcal{L}_{LM}
\ee
with the explicit expression
\be
\mathcal{L}_{LM}
=
\mathcal{L}_{LM}[\theta^\alpha, \omega^{\alpha}_{\fant{a}\beta}, \lambda^\alpha, \Theta^\alpha]
=
\lambda^\alpha\wdg \Theta_\alpha
\ee
constraining the independent metric compatible connection to be Levi-Civita. Then the field equations for pseudo-Riemannian model that follow from the Lagrangian 4-form in (1) is explicitly recovered from the constrained first order formalism as follows.

First, by making use of the variational derivative expressions
 \be
 \delta\Theta^\alpha
 =
 D\delta\theta^\alpha+\delta\omega^{\alpha}_{\fant{a}\beta}\wdg \theta^\beta
 \ee
 and
 $\delta\Omega^{\alpha}_{\fant{a}\beta}
 =
 D\delta\omega^{\alpha}_{\fant{a}\beta},
 $ as well, one finds that  the total variational derivative of extended Lagrangian density with respect to the gravitational variables has  the general form
\be\label{gen-variational-der}
\delta\mathcal{L}_e
=
\delta\omega_{\alpha\beta}\wdg\left[D\frac{\partial \mathcal{L}_g}{\partial \Omega^{\alpha\beta}}
-
\frac{1}{2}\left(
\theta^\alpha\wdg \frac{\partial \mathcal{L}_{LM}}{\partial\Theta^\beta}
-
\theta^\beta
\wdg \frac{\partial \mathcal{L}_{LM}}{\partial\Theta^\alpha}\right)\right]
+
\delta\theta^{\alpha}
\wdg
\left(\frac{\partial \mathcal{L}_g}{\partial\theta^{\alpha}}+\frac{\partial \mathcal{L}_m}{\partial\theta^{\alpha}}+D\frac{\partial \mathcal{L}_{LM}}{\partial\Theta^{\alpha}}\right)
+
\delta\lambda^\alpha\wdg \Theta_\alpha
\ee
up to a disregarded boundary term.
The partial derivative of the Lagrangian form with respect to a $p$-forms  is very convenient and an expression for it can be  obtained from the corresponding expressions for the variational derivative. Although it is not necessary (and also neither helpful nor practical) in what follows, it is possible to relate the partial derivatives with respect  to forms above to the ordinary partial derivatives with respect to the associated tensor components \cite{kopczynsky}.

The general formula (\ref{gen-variational-der}) is to be applied to the QC gravity Lagrangian density (\ref{extended-lag-def}). To this end,
the following definitions of auxiliary tensor-valued forms are convenient for the study of the corresponding field equations.
In particular, it is convenient to introduce the auxiliary 2-form $X^{\alpha\beta}$ for the partial derivatives
of the gravitational  part with respect to curvature 2-forms as
\be\label{4d-PI-expression}
\frac{\partial \mathcal{L}_g}{\partial \Omega^{\alpha\beta}}
\equiv
\tfrac{1}{2}*\left(X^{\alpha\beta}+\kappa^{-2}\theta^{\alpha\beta}\right)
\ee
and it is possible to show that, for the QC terms in the Lagrangian (\ref{lag-def}), the auxiliary 2-form $X^{\alpha\beta}$ are explicitly given by the following expression
\be\label{aux-X-def}
X^{\alpha\beta}
=
-\tfrac{1}{2}\left[\theta^{\alpha}\wdg(bR^\beta+aR\theta^\beta)-\theta^{\beta}\wdg(bR^\alpha+aR\theta^\alpha)\right]
\ee
which is already in a suitable form for the projection of the field equations  to a null coframe.
The auxiliary 2-form $X^{\alpha\beta}$ can be written as a linear sum of  self-dual and anti-self-dual parts.

Yet another convenient and useful  auxiliary 3-form  is defined by
\be
\Pi^{\alpha\beta}\equiv D*X^{\alpha\beta}+\omega^{\alpha}_{\fant{a}\mu}\wdg *X^{\mu\beta}+\omega^{\beta}_{\fant{a}\mu}\wdg *X^{\alpha\mu}
\ee
$\Pi^{\alpha\beta}=-\Pi^{\beta\alpha}$ are related to the connection equations to be discussed below. In the context of metric-affine gravity \cite{hehl-mccrea-mielke-neemann}, the partial derivatives corresponding to the auxiliary forms $*X^{\alpha\beta}$ and $\Pi^{\alpha\beta}$ are called the gravitational gauge field momenta in the context of metric-affine gravity, see also \cite{obukhov-wave}.  Moreover,  the auxiliary 2-forms $X^{\alpha\beta}$ also appear in the explicit expression for variational derivative of the gravitational part with respect to the basis coframe 1-forms. Explicitly, one finds the  following partial derivative
\be\label{coframe-partial-der}
\frac{\partial \mathcal{L}_g}{\partial \theta^{\alpha}}
=
\frac{1}{\kappa^2}(-*G^\alpha
+
\Lambda_0*\theta^\alpha)
+
*T^\alpha[\Omega^{\mu}_{\fant{a}\nu},X^{\mu\nu}]
\ee
where, the last term on the right-hand side  is the derivatives of the QC part with respect to the coframe 1-forms and it can be defined by
\be
*T^\alpha[\Omega^{\mu}_{\fant{a}\nu},X^{\mu\nu}]
\equiv
-
\tfrac{1}{4}(i^\alpha\Omega_{\mu\nu})\wdg *X^{\mu\nu}
+
\tfrac{1}{4}\Omega_{\mu\nu}\wdg i^\alpha*X^{\mu\nu}
\ee
for the sake of brevity of the expression (\ref{coframe-partial-der}). Similarly,  the energy-momentum 3-form
\be\label{en-mom-F}
\frac{\partial\mathcal{L}_m}{\partial\theta^\alpha}
\equiv
*\tau^\alpha[F]
=
\tfrac{1}{2}[(i^\alpha F)\wdg *F-F\wdg i^\alpha *F]
\ee
energy -momentum 3-form of the Maxwell field which is the  the only matter field considered. It is easy to write Maxwell's equations and the electromagnetic energy-momentum form in terms of the self-dual Faraday 2-form $\mathcal{F}\equiv\frac{1}{2}(F+i*F)$ relative to a null coframe as well  (see Eq. (\ref{en-mom-F-2-form}) in the Appendix below).

In terms of the definitions above, the coframe (metric) field equations then take the form ${\delta\mathcal{L}_e}/{\delta\theta^\alpha}\equiv *E^\alpha=0$
in terms of the vector-valued 1-form $E^\alpha=E^{\alpha}_{\fant{a}\beta}\theta^\beta$ with
\be\label{general-metric-eqns}
*E^\alpha
=
\frac{1}{\kappa^2}(-*G^\alpha
+
\Lambda_0*\theta^\alpha)
+
D\lambda^\alpha
+
*T^\alpha[\Omega^{\mu}_{\fant{a}\nu}, X^{\mu\nu}]
+
*\tau^\alpha[F].
\ee
The metric equations (\ref{general-metric-eqns}) are to be supplemented with the matter field equations, namely the source-free Maxwell equations, $dF=d*F=0$.

Even though a set of a basis local orthonormal coframe is adopted in the derivation of the metric equations (\ref{general-metric-eqns}) derived in \cite{baykal}, the projection of the equations to a null coframe  follows  simply  by specializing the numerical indices to a null coframe. In fact, in the form given above they allow one to introduce a real null coframe \cite{gps-frame} or  even a hybrid  rigid coframe with half-null  and half-orthonormal basis 1-forms used for example in a study of plane-fronted waves in arbitrary dimensions \cite{obukhov-wave,malek-pravda}.
Moreover, although the QC field equations $*E^\alpha=0$ relative to a null coframe are still somewhat lengthy to be written out  explicitly even in the tensorial form, the  field equations are such that they allow the auxiliary tensor fields, for example $X^{\alpha\beta}$ defined above,  to be  expressed in terms of the spinor components in a suitable form. Consequently, if the components of the differential forms in (\ref{general-metric-eqns}) can be expressed in terms of the spinor definitions  with the help of the formulas given in Appendix, then (\ref{general-metric-eqns}) yields  NP-type complex scalar equations in terms of spinors.

Returning to the variational derivative with respect to connection 1-form, where it assumed that it is involved only in the gravitational sector if one assumes the minimal coupling of matter and vanishing torsion. The connection equations lead to an algebraic equation for the forms $\lambda^\alpha$. They  can be written out as
\be\label{pi-general-formulae}
\Pi^{\alpha\beta}
=
\tfrac{1}{2}(\theta^\alpha\wdg\lambda^\beta-\theta^\beta\wdg \lambda^\alpha).
\ee
The right-hand side of (\ref{pi-general-formulae}) is to be calculated  by imposing the constraint vanishing torsion constraint $\Theta^\alpha=0$. Note at this point that in four dimensions, $\lambda^\alpha$ is a vector-valued $2$-form whereas $\Pi^{\alpha\beta}$ is an antisymmetric (second rank) tensor-valued $3$-form
and they  both have, at most, 24 independent components. The   equivalence of the tensor-valued forms implies that the algebraic relation (\ref{pi-general-formulae}) can uniquely be inverted to have
\be\label{general-expression-lag-mult}
\lambda^\alpha
=
2i_\beta\Pi^{\beta\alpha}+\tfrac{1}{2}\theta^\alpha\wdg i_\mu i_\nu \Pi^{\mu\nu}.
\ee
The expression on the right-hand side can be obtained by calculating two successive contractions of (\ref{pi-general-formulae}).
The properties  of the tensor-valued $(n-1)$ form $\Pi^{\alpha\beta}$ is crucial in the study of Lagrangian densities below in the sense that it also determines  the properties of the fourth-order terms that contributes to the metric equations.
Consequently, the $\lambda^\alpha$ in QC equations (\ref{general-metric-eqns}) explicitly takes the form
 \be\label{lag-mult-general-form}
\lambda^\alpha
=
-*D[bR^\alpha+(2a+\tfrac{1}{2}b)R\theta^\alpha]
 \ee
by inserting the expression for $X^{\alpha\beta}$ from (\ref{aux-X-def}) into (\ref{general-expression-lag-mult}). Moreover, as  is well known,  the contribution of the Einstein-Hilbert part to $\lambda^\alpha$ vanishes identically by imposing the vanishing torsion constraint.

Eventually, one obtains the metric equations  in a form such that all the fourth-order terms are contained in the Lagrange multiplier term  which are linear in the curvature and the remaining terms are  quadratic in curvature components. Finally, variational derivative with respect to Lagrange multiplier $\lambda^\alpha$ yields the equation $\Theta^\alpha=0$ and the connection and coframe equations are evaluated with the constraint $\Theta^\alpha=0$. For a given quadratic curvature Lagrangian in four dimensions by assuming particular values for the coupling constants $a$ and $b$, it is possible  to simplify  the field equations further by making use of the identities that curvature tensor satisfies. Finally, note that although the most of the following work is confined to four dimensions, the QC field equations, in fact, holds in arbitrary dimensions $n\geq3$ \cite{baykal,baykal-delice}.

The trace of the metric equations can be found by wedging Eq. (\ref{general-metric-eqns}) with $\theta_{\alpha}$ and summing over the free index. In four dimensions, the quadratic curvature part, namely $*T^\alpha$, does  not contribute to the trace $E^\alpha_{\phantom{a}\alpha}\equiv E$. The trace defined by $\theta_\alpha\wdg *E^\alpha\equiv E*1$ explicitly takes the form
\be
E*1
=
\frac{1}{\kappa^2}(R*1
+
4\Lambda_0*1)
+
D*i_\alpha D[bR^\alpha+(2a+\tfrac{1}{2}b)R\theta^\alpha]=0
\ee
involving only the terms linear in the curvature components. This peculiar property of the trace for the QC part of the Lagrangian is unique  to four dimensions see, for example \cite{baykal}. The trace expression can further be simplified by using the contracted second Bianchi identity,   which can be written conveniently  as  $D*G^\alpha=0$ where $G^\alpha$ is the Einstein 1-form $G^\alpha\equiv G^\alpha_{\fant{a}\beta}\theta^\beta\ =R^\alpha-\tfrac{1}{2}R\theta^\alpha$.
Consequently, the left-hand side of the trace expression simplifies to
\be
\frac{1}{\kappa^2}(R*1
+
4\Lambda_0*1)
+
(3a+b)d*dR=0.
\ee
Furthermore, for the QC coupling parameters satisfying $3a+b=0$, it is well known that the QC part of the Lagrangian (\ref{lag-def}) is equivalent  to conformally invariant Weyl gravity and, accordingly, the QC part does not contribute to the trace in this case.

As a final remark on the structure of  the general QC equations and the auxiliary forms introduced above in comparison to the Einstein-Hilbert Lagrangian $\mathcal{L}_{EH}$, and the Einstein field equations, note that it is possible to rewrite the Einstein-Hilbert Lagrangian form  as
\begin{align}
\mathcal{L}_{EH}
&=
-
\tfrac{1}{2}\Omega_{\alpha\beta}\wedge*\theta^{\alpha\beta}
\nonumber\\
&=
\tfrac{1}{2}d\theta^{\alpha}\wedge*F_{\alpha}
+
d\left(\theta_{\alpha}\wedge*d\theta^{\alpha}\right)
\nonumber\\
&=
\tfrac{1}{2}\left(d\theta_{\alpha}\wedge \theta_{\beta}\right)
\wedge*\left(d\theta^{\beta}\wedge \theta^{\alpha}\right)
-
\tfrac{1}{4}\left(d\theta^{\alpha}\wedge \theta_{\alpha}\right)
\wedge*\left(d\theta^{\beta}\wedge \theta_{\beta}\right)
+
d\left(\theta_{\alpha}\wedge*d\theta^{\alpha}\right)\label{einstein-hilbert-qc-form}
\end{align}
(see, \cite{straumann,thirring} for more details).
The third line above follows from the definition the familiar Sparling-Thirring 2-form $*F^{\alpha}$ as
\begin{align}
*F^{\alpha}
&\equiv
-\tfrac{1}{2}\omega_{\mu\nu}\wdg *\theta^{\alpha\mu\nu}
\nonumber\\
&=
\theta_\beta\wdg *
(d\theta^{\beta}\wedge \theta^{\alpha})-\tfrac{1}{2}\theta^\alpha\wdg *(d\theta^{\beta}\wedge \theta_{\beta}).
\end{align}
The Einstein-Hilbert Lagrangian in (\ref{einstein-hilbert-qc-form})
is explicitly in a form that is quadratic in the variable $d\theta^\alpha\wdg \theta_\beta$. Accordingly,  by eliminating the Levi-Civita connection 1-forms in favor of $d\theta^\alpha$ and its contractions, the Einstein vacuum field equations
can be written in a Maxwell-like form as
\be
d*F^\alpha-(i_\alpha  d\theta^\beta) \wdg *F_\beta+d\theta^\beta \wdg i_\alpha *F_\beta
=
0
\ee
in terms of the variables $\{\theta^\alpha\}$ and   exterior derivative $\{d\theta^\alpha\}$.
This form of the vacuum Einstein field equations has  formal structural similarity with the generic QC vacuum field equations when they are written in terms of
the auxiliary forms in the form
\be\label{pure-qc-eqn}
-D*D[bR^\alpha+(2a+\tfrac{1}{2}b)R\theta^\alpha]
-
\tfrac{1}{4}(i^\alpha\Omega_{\mu\nu})\wdg *X^{\mu\nu}
+
\tfrac{1}{4}\Omega_{\mu\nu}\wdg i^\alpha*X^{\mu\nu}
=0.
\ee
Note however that Eqs. (\ref{einstein-hilbert-qc-form}) are second-order in the derivatives of $\theta^\alpha$ (equivalently, in the metric components) whereas
the field equations (\ref{pure-qc-eqn}) are second order derivatives of the curvature components.

An important feature of the field equations (\ref{general-metric-eqns}) and (\ref{lag-mult-general-form}) for the discussion below is that they are valid relative to any rigid coframe  and thus they provide a convenienience for calculational scheme  relative to a null coframe. In particular, the auxiliary forms of the QC gravity introduced above are suitable for this purpose.

\subsection{QC field equations relative to a null coframe}

In order to write down the field equations for the QC model (\ref{lag-def}) relative to a NP null coframe, one has to identify  the tensorial components in terms of the complex spinor scalars (the spin coefficients and the curvature spinors etc.). Therefore, it is convenient and is advantageous to adopt a NP null coframe conventions similar to those \cite{exact-sol-stephani}, see also \cite{guven-yoruk}. Although the expressions for QC field equations are also lengthy relative to a null coframe,  the formulas and  definitions for the field equations for the QC gravity above allow one to introduce the NP spinor quantities in terms of convenient auxiliary tensor-valued forms.

All the numerical indices below refer exclusively to the null coframe.
Relative to a NP null coframe description of the geometrical quantities in terms of differential forms given in the Appendix, it is convenient to express the auxiliary 2-forms $X^{\alpha\beta}$  in terms of the complex traceless-Ricci spinors $\Phi_{ik}$ and the scalar curvature $R$ as
\be\label{aux-X-spin-defs}
\begin{split}
X^{0}_{\fant{0}3}
&=
\tfrac{1}{2}b[-\Phi_{00}n\wdg m-\Phi_{01}(l\wdg  n+m\wdg \bar{m})+\Phi_{02}l\wdg \bar{m}]
+
\tfrac{1}{12} (a+\tfrac{1}{4}b)Rl\wdg m
\\
X^{1}_{\fant{0}2}
&=
\tfrac{1}{2}b[+\Phi_{20}n\wdg m
+
\Phi_{21}(l\wdg n+m\wdg\bar{m})
-
\Phi_{22}l\wdg \bar{m}
]
+
\tfrac{1}{12}(a+\tfrac{1}{4}b)Rn\wdg \bar{m}
\\
\tfrac{1}{2}(X^{0}_{\fant{0}0}-X^{3}_{\fant{0}3})
&=
\tfrac{1}{2}b[
+
\Phi_{10}n\wdg m
+
\Phi_{11}(l\wdg n+m\wdg\bar{m})
-
\Phi_{12}l\wdg \bar{m}
]
-
\tfrac{1}{24}(a+\tfrac{1}{4}b)R(l\wdg n-m\wdg\bar{m}).
\end{split}
\ee

These expressions  can readily be obtained by making use of the curvature spinor definitions provided in the Appendix
and specializing (\ref{aux-X-def}) to a NP null coframe. There are six number of real and independent auxiliary form $X^{\alpha\beta}$ and the remaining complex three can be obtained by complex conjugation of the ones above. The expressions (\ref{aux-X-spin-defs}) then can be used to express
$*T^\alpha[\Omega^{\mu}_{\fant{a}\nu}, X^{\mu\nu}]$ in terms of Ricci spinors and the scalar curvature relative to a null coframe.
Note that $X^{\alpha\beta}$ has anti-selfdual terms, namely the terms in square brackets on the right-hand side   and scalar curvature terms are self-dual.
The expressions for the auxiliary 2-form $X^{\alpha\beta}$ is to be used in the expression for auxiliary form $T^\alpha[\Omega^{\mu}_{\fant{a}\nu}, X^{\mu\nu}]$
\be
\begin{split}
*T^0[\Omega^{\mu}_{\fant{a}\nu}, X^{\mu\nu}]
&=
-
\tfrac{1}{4}(i_{l^\sharp}\Omega_{\mu\nu})\wdg *X^{\mu\nu}
+
\tfrac{1}{4}\Omega_{\mu\nu}\wdg i_{l^\sharp}*X^{\mu\nu}
\\
*T^1[\Omega^{\mu}_{\fant{a}\nu}, X^{\mu\nu}]
&=
-
\tfrac{1}{4}(i_{n^\sharp}\Omega_{\mu\nu})\wdg *X^{\mu\nu}
+
\tfrac{1}{4}\Omega_{\mu\nu}\wdg i_{n^\sharp}*X^{\mu\nu}
\\
*T^2[\Omega^{\mu}_{\fant{a}\nu}, X^{\mu\nu}]
&=
+
\tfrac{1}{4}(i_{m^\sharp}\Omega_{\mu\nu})\wdg *X^{\mu\nu}
-
\tfrac{1}{4}\Omega_{\mu\nu}\wdg i_{m^\sharp}*X^{\mu\nu}
\end{split}
\ee
and by definition the third components can be obtained from the second one by complex  conjugation $\bar{T}^2[\Omega^{\mu}_{\fant{a}\nu}, X^{\mu\nu}]=T^3[\Omega^{\mu}_{\fant{a}\nu}, X^{\mu\nu}]$.

The covariant exterior derivative of the auxiliary form $X^{\alpha\beta}$ also appears in the fourth-order part of the metric field equations.
By making use of the antisymmetry properties  $\omega_{\alpha\beta}+\omega_{\beta\alpha}=0$ and $X_{\alpha\beta}+X_{\beta\alpha}=0$, one can find
the following expression for covariant derivative of $D*X^{\alpha}_{\fant{a}\beta}\equiv\Pi^{\alpha}_{\fant{a}\beta}$
\be
\begin{split}
D*X^{0}_{\fant{a}3}
&=
d*X^{0}_{\fant{a}3}
-
\omega^{0}_{\fant{a}3}\wdg*(X^{0}_{\fant{a}0}-X^{3}_{\fant{a}3})
+
(\omega^{0}_{\fant{a}0}-\omega^{3}_{\fant{a}3})\wdg*X^{0}_{\fant{a}3}
\\
D*X^{1}_{\fant{a}2}
&=
d*X^{1}_{\fant{a}2}
+
\omega^{1}_{\fant{a}2}\wdg*(X^{0}_{\fant{a}0}-X^{3}_{\fant{a}3})
-
(\omega^{0}_{\fant{a}0}-\omega^{3}_{\fant{a}3})\wdg*X^{1}_{\fant{a}2}
\\
D*(X^{0}_{\fant{a}0}-X^{3}_{\fant{a}3})
&=
d*(X^{0}_{\fant{a}0}-X^{3}_{\fant{a}3})
+
2\omega^{0}_{\fant{a}3}\wdg*X^{1}_{\fant{a}2}
-
2\omega^{1}_{\fant{a}2}\wdg*X^{0}_{\fant{a}3}.
\end{split}
\ee
These equations can be written out in terms of spin coefficients with the help of the expressions (\ref{spinor-def}).
In consequence, by calculating the contraction of the expressions above one can calculate  the components of the Lagrange multiplier form $\lambda^\alpha$
with the help of the general formula (\ref{general-expression-lag-mult}) which read
\be
\begin{split}
\lambda^0
&=
-2i_{l^\sharp}\Pi^{01}+2i_{m^\sharp}\Pi^{03} +2i_{\bar{m}^\sharp}\Pi^{02}
\\
&\phantom{px}+
l\wdg
\left(
i_{l^\sharp} i_{n^\sharp}\Pi^{10}
-
i_{l^\sharp} i_{m^\sharp}\Pi^{13}
-
i_{l^\sharp} i_{\bar{m}^\sharp}\Pi^{12}
-
i_{n^\sharp} i_{m^\sharp}\Pi^{03}
-
i_{n^\sharp} i_{\bar{m}^\sharp}\Pi^{02}
+
i_{m^\sharp} i_{\bar{m}^\sharp}\Pi^{23}
\right)
\\
\lambda^1
&=
+
2i_{n^\sharp}\Pi^{01}+2i_{m^\sharp}\Pi^{13}+2i_{\bar{m}^\sharp}\Pi^{12}
\\
&\phantom{px}+
n\wdg
\left(
i_{l^\sharp} i_{n^\sharp}\Pi^{10}
-
i_{l^\sharp} i_{m^\sharp}\Pi^{13}
-
i_{l^\sharp} i_{\bar{m}^\sharp}\Pi^{12}
-
i_{n^\sharp} i_{m^\sharp}\Pi^{03}
-
i_{n^\sharp} i_{\bar{m}^\sharp}\Pi^{02}
+
i_{m^\sharp} i_{\bar{m}^\sharp}\Pi^{23}
\right)
\\
\lambda^2
&=
+
2i_{n^\sharp}\Pi^{02}+2i_{l^\sharp}\Pi^{12}-2i_{{m}^\sharp}\Pi^{23}
\\
&\phantom{px}+
m\wdg
\left(
i_{l^\sharp} i_{n^\sharp}\Pi^{10}
-
i_{l^\sharp} i_{m^\sharp}\Pi^{13}
-
i_{l^\sharp} i_{\bar{m}^\sharp}\Pi^{12}
-
i_{n^\sharp} i_{m^\sharp}\Pi^{03}
-
i_{n^\sharp} i_{\bar{m}^\sharp}\Pi^{02}
+
i_{m^\sharp} i_{\bar{m}^\sharp}\Pi^{23}
\right)
\end{split}
\ee
with $\lambda^3=\bar{\lambda}^2$. These general expression has sufficient generality to calculate $\lambda^\alpha$ for a given Lagrangian. For the QC Lagrangian  above,  the  explicit form  of the Lagrange multiplier 2-forms then take the form
\be\label{null-comps-lag-multiplier}
\begin{split}
\lambda^0
&=
*[d(bR^0+(2a+\tfrac{1}{2}b)Rl)
+
\omega^{0}_{\fant{a}0}\wdg (bR^0+(2a+\tfrac{1}{2}b)Rl)
\\
&\phantom{px}+
\bar{\omega}^{0}_{\fant{a}3}\wdg (bR^2+(2a+\tfrac{1}{2}b)Rm)
+
\omega^{0}_{\fant{a}3}\wdg (bR^3+(2a+\tfrac{1}{2}b)R\bar{m})]
\\
\lambda^1
&=
*[d(bR^1+(2a+\tfrac{1}{2}b)Rn)
-
\omega^{0}_{\fant{a}0}\wdg (bR^1+(2a+\tfrac{1}{2}b)Rn)
\\
&\phantom{px}+
\omega^{1}_{\fant{a}2}\wdg (bR^2+(2a+\tfrac{1}{2}b)Rm)
+
\bar{\omega}^{1}_{\fant{a}2}\wdg (bR^3+(2a+\tfrac{1}{2}b)R\bar{m})]
\\
\lambda^2
&=
*[d(bR^2+(2a+\tfrac{1}{2}b)Rm)
-
\omega^{3}_{\fant{a}3}\wdg (bR^2+(2a+\tfrac{1}{2}b)Rm)
\\
&\phantom{px}+
\bar{\omega}^{1}_{\fant{a}2}\wdg (bR^0+(2a+\tfrac{1}{2}b)Rl)
+
{\omega}^{0}_{\fant{a}3}\wdg (bR^1+(2a+\tfrac{1}{2}b)Rn)]
\end{split}
\ee
which can be obtained more directly by making use of (\ref{lag-mult-general-form}) as well.

Finally, the expressions obtained by the above formulas for the Lagrange multiplier forms are to be inserted into the  expression of for the  covariant derivative of the Lagrange multiplier and the covariant exterior derivatives of a vector valued $2$-form $\lambda^\alpha$ relative to null coframe explicitly read
\be\label{cov-der-lamda-null-cof}
\begin{split}
D\lambda^0
&=
d\lambda^0+\omega^{0}_{\fant{a}0}\wdg \lambda^0+\bar{\omega}^{0}_{\fant{a}3}\wdg \lambda^2+\omega^{0}_{\fant{a}3}\wdg \lambda^3
\\
D\lambda^1
&=
d\lambda^1-\omega^{0}_{\fant{a}0}\wdg \lambda^1+{\omega}^{1}_{\fant{a}2}\wdg \lambda^2+\bar{\omega}^{1}_{\fant{a}2}\wdg \lambda^3
\\
D\lambda^2
&=
d\lambda^2+\bar{\omega}^{1}_{\fant{a}2}\wdg \lambda^0+\omega^{0}_{\fant{a}3}\wdg \lambda^1-\omega^{3}_{\fant{a}3}\wdg \lambda^2
\end{split}
\ee
and that $D\lambda^3=D\bar{\lambda}^2$.
Consequently, with the help of (\ref{aux-X-spin-defs})-(\ref{cov-der-lamda-null-cof}) and also using spinor components of the Ricci 1-form $R^\alpha$ using (\ref{general-einstein-3forms}), the expression (\ref{general-metric-eqns}) for $*E^\alpha$ can straightforwardly be projected to a  NP null coframe in full generality and the Lagrange multiplier terms can be calculated in stages by using  (\ref{null-comps-lag-multiplier}) in (\ref{cov-der-lamda-null-cof}).

In the general case, (\ref{cov-der-lamda-null-cof}) eventually yields a set of  equations involving the second-order derivatives of the curvature spinors.
Therefore, in QC gravity, instead of determining the Ricci spinors algebraically via Einstein field equations, $*G^\alpha=\kappa^2*\tau^\alpha$, and subsequently to use them in Ricci identities, the QC field equations yield second-order equations for curvature spinors which cannot be just inserted into the Ricci identities. However, the use of the NP spin coefficient formalism may still provide a convenient simplifying scheme of calculations in a study of exact solutions to the QC general gravity. Although the QC field equations $*E^\alpha=0$ yield equations for second-order derivatives of the Ricci spinors, a NP null coframe formulation may still provide a convenient scheme for an investigation of  algebraically special solutions to the general QC gravity in a manageable form compared to the tensorial methods.

The integrability conditions for the Cartan's second structure equations,  or the Bianchi identities, $D\Omega^{\alpha}_{\fant{a}\beta}=0$, relative to a real orthonormal coframe can be considered as the set of equations for 3-forms and component-wise they add up to a total of 24  equations. Relative to  a complex null coframe,   on the other hand, the number of Bianchi identities is reduced to 12 by making use of the complex curvature 2-forms $\Omega^{0}_{\fant{a}3}, \Omega^{1}_{\fant{a}2}$ and $\frac{1}{2}(\Omega^{0}_{\fant{a}0}-\Omega^{3}_{\fant{a}3})$. Thus, Cartan's second structure equations for each of the complex curvature 2-form leads to four scalar equations. In their seminal work\cite{newman-penrose},  Newman and Penrose  introduced  eight complex scalar equations for the Bianchi identity, all of which can explicitly be derived from the components of the three form equations $D\Omega^{0}_{\fant{a}3}=0$ and $D\Omega^{1}_{\fant{a}2}=0$. The scalar equations for the second Bianchi identity in the literature is usually given as a set of eleven number of complex scalar equations \cite{exact-sol-stephani, chandra}. Out of total eleven, eight scalar equations can be derived from the components of the three form equations $D\Omega^{0}_{\fant{a}3}=D\Omega^{1}_{\fant{a}2}=0$ while the remaining three scalar  equations follow from the contracted Bianchi identity. More precisely, instead of  the equations that follow from $D(\Omega^{0}_{\fant{a}0}-\Omega^{3}_{\fant{a}3})=0$,  the scalar equations that follow from $D*G^0=0$, $D*G^1=0$ and $D*G^2=D*\bar{G}^3=0$ are included in scalar equations of the Bianchi identity \cite{chandra}.
The Bianchi identities, in the complex scalar form, involve first-order derivatives of the
curvature spinors.  However, it is well known that contracted Bianchi identities, $D*G^\alpha=d*G^\alpha+\omega^{\alpha}_{\fant{a}\beta}\wdg *G^\beta=0$, are independent of the field equations as they follow from the diffeomorphism invariance of the Einstein-Hilbert action.

In the light of these remarks, returning now to  the QC gravity model (\ref{lag-def}) above, the identities resulting from the diffeomorphism invariance now becomes $D*E^\alpha=d*E^\alpha+\omega^{\alpha}_{\fant{a}\beta}\wdg *E^\beta=0$ \cite{kopczynsky} and these equations involve the third-order derivatives of the curvature spinors in addition to the Bianchi identities $D\Omega^{\alpha}_{\fant{a}\beta}=0$.

Since  null coframe formulation provides a  quite efficient framework in the study of exact solutions and in particular the radiative metrics  in the  general theory of relativity, one can expect that NP-like null coframe formulation of the QC model to have an analogous potential efficiency.
It may also find  application,  for example, in obtaining   perturbative gravitational wave solutions of the QC gravity   as discussed in \cite{neto-perturbative} in terms of an expansion in the QC coupling constants.

In the following section, the above formulas  are applied,  as an application of the above null coframe formulation of the QC field equations, to study a particular impulsive wave solutions propagating in algebraically special backgrounds of Petrov type D. The discussion in the following two sections can be considered as extensions of the previous exact  solutions reported in \cite{ortaggio} to general QC gravity.

\section{An application of the formalism}

A relatively simple application the formalism provided above in terms of exterior algebra and auxiliary tensor-valued forms is the discussion of the   wave solutions dealing with metrics that are slightly more complicated then the usual $pp$-wave metrics. Before introducing a new family of gravitational wave solutions  on algebraically special background spacetimes that are all of the direct product of the form $M_1\times M_2$ with the further assumption that both $M_1$ and $M_2$ are  constant curvature spacetimes, it is first  convenient to show that these are actually the solutions of QC gravity models provided that a system of algebraic equations for the parameters of such two-spaces $M_1$ and $M_2$ are satisfied.

\subsection{Direct product solutions}

In general, for a study of a radiative spacetime metric it is both natural and advantageous to adopt the NP spin coefficient formalism in one form or another. In  the present framework, this is achieved by simply introducing a null coframe and formulate the field equations derived above with respect to a null coframe. It is also desirable to relate the null coframe components of tensors and forms  to the spinor quantities in the NP spin coefficient formalism. This is  provided in the Appendix.
Recently, in parallel to the work in this section,  a new family of exact solutions to QC gravity in five dimensions  in the form of direct product of two spaces have been introduced in \cite{clement}.

The background metrics  that will be considered for impulsive gravitational waves will be assumed to be of the direct product of two two-dimensional manifolds $M_1\times M_2$ of the form
\be\label{background-metrics}
g=
\frac{1}{\Omega^2}(du\ot dv+dv\ot du)
-
\frac{1}{P^2}(d\zeta\ot d\bar{\zeta}+d\bar{\zeta}\ot d{\zeta})
\ee
where the (real) conformal factors of the two dimensional parts are assumed  to be of the form
\be
\Omega(u,v)=1-\frac{k_1}{2l_1^2}uv, \qquad  P(\zeta,\bar{\zeta})=1+\frac{k_2}{2l_2^2}\zeta\bar{\zeta}
\ee
respectively. $u, v$ are  real null coordinates on $M_1$ with Lorentzian signature whereas $\zeta$ is complex spacelike coordinate on $M_2$ with Euclidean signature. The constants $k_1$ and $k_2$ take the values $0, \mp1$ and $l_1$ and $l_2$ are constants related to constant curvatures of the two dimensional spaces $M_1$ and $M_2$ respectively. These type D spacetimes  \cite{bertotti,robinson,plebansky-hacyan,nariai}  are listed in Table I for convenience and for further properties of these spacetimes,  the reader is referred to \cite{griffiths-podolsky-exact-sol,ortaggio-podolsky}.

\begin{table}
\caption{Possible direct product spacetimes $M_1\times M_2 $with $l_1=l_2$. See, Chapter 7 in \cite{griffiths-podolsky-exact-sol} or \cite{ortaggio-podolsky}.}
\begin{center}
\begin{tabular}{c||c|c|c|c|c}
  \hline
  \emph{Spacetime} &\quad $M_1\times M_2$\quad & $k_1$ & $k_2$ & $\Phi_{11}$ & $\Lambda_0$ \\
  \hline\hline
Minkowski& $\mathbb{R}^{(1,1)}\times \mathbb{R}^{2}$& 0& 0& =0&=0\\
\hline
Nariai&$dS_2\times S_2$&$+1$&$+1$& =0 & $>0$\\
\hline
anti-Nariai&$AdS_2\times H_2$&$-1$&$-1$& =0 & $<0$\\
\hline
Bertotti-Robinson&$AdS_2\times S_2$&$-1$ & $+1$ & $>0$ & $=0$\\
\hline
\multirow{2}{*}{Pleba\`{n}ski-Hacyan}&$\mathbb{R}^{(1,1)}\times S_2$ & 0 &$+1$ & $>0$&$>0$ \\\cline{2-6}
                                     &$AdS_2\times\mathbb{R}^{2}$ &$-1$&0&$>0$&$<0$\\
\hline

\end{tabular}
\end{center}
\end{table}

In terms of   local null coordinates $\{x^a\}=\{u, v, \zeta, \bar{\zeta}\}$, it is natural to define the local null coframe basis 1-forms $\{\theta^\alpha\}$ as
\be
l=\frac{1}{\Omega}du,\qquad n=\frac{1}{\Omega}dv,\qquad m=\frac{1}{P}d\zeta, \qquad \bar{m}=\frac{1}{P}d\bar{\zeta}
\ee
for $\alpha=0,1,2,3$  respectively and the metric then assumes the  standard form
\be
g=
l\ot n+n\ot l-m\ot\bar{m}-\bar{m}\ot m.
\ee
$l$ is the repeated principle null eigen 1-form.
The class of distribution-valued metrics of the form (\ref{background-metrics}) has previously been introduced by constraining six dimensional $pp$-wave metrics \cite{ortaggio-podolsky}.
The associated set of null frame fields $\{e_\alpha\}=\{n^\sharp, l^\sharp, -\bar{m}^\sharp, -m^\sharp\}$ are
\be
\Delta
=\Omega(\partial_u-H\partial_v), \qquad D=\Omega\partial_v,\qquad \bar{\delta}=-P\partial_{{\zeta}}, \qquad {\delta}=-P\partial_{\bar{\zeta}}
\ee
for $\alpha=0,1,2,3$ respectively. In terms of the null coframe 1-forms, the exterior derivative operator has the expansion
\be
d
=
l\Omega\partial_u+n\Omega\partial_v+Pm\partial_\zeta+P\bar{m}\partial_{\bar{\zeta}}
\ee
acting on scalars.
Using the exteriors derivatives of the basis coframe 1-forms relative to the null coframe, by specializing indices of the Cartan's first structure equations
\be\label{cartan-1eqns-general}
d\theta^\alpha+\omega^{\alpha}_{\fant{a}\beta}\wdg \theta^\beta=0
\ee
to the null coframe, and taking the metric compatibility relation for the connection 1-forms, $\eta_{\alpha\mu}\omega^{\mu}_{\fant{a}\beta}+\eta_{\beta\mu}\omega^{\mu}_{\fant{a}\alpha}=0$ into account, (\ref{cartan-1eqns-general}) can be solved for  the Levi-Civita connection 1-forms $\omega^{\alpha}_{\fant{a}\beta}$ as
\be
\omega^{0}_{\fant{a}0}
=
\frac{k_1}{2l_1^2}(vl-un),\qquad
\omega^{3}_{\fant{a}3}
=
\frac{k_2}{2l_2^2}(\bar{\zeta}m-\zeta\bar{m})
\ee
with all other components vanishing (see also the general formulas for the connection 1-forms in the Appendix).
  All the other connection 1-form can be obtained from (\ref{connection-forms}) by complex conjugation
or else by the metric compatibility relation  $\eta_{\alpha\mu}\omega^{\mu}_{\fant{a}\beta}+\eta_{\beta\mu}\omega^{\mu}_{\fant{a}\alpha}=0$
relative to the null coframe. The nonvanishing spin coefficients relative to the null coframe adopted then have the following expressions
\be
\gamma
=-\frac{k_2}{2l_1^2}v,
\qquad
\alpha
=
-\bar{\beta}=\frac{k_2}{2l_2^2}\bar{\zeta}
\ee

The corresponding curvature spinors and the null coframe components of the curvature 2-form can be calculated by using the Cartan's  second structure
equations (\ref{cartan-SE2}) which yield the following expressions
\be\label{product-sym-curvature-forms}
\begin{split}
\Omega^{0}_{\fant{a}3}
&=
0
\\
\Omega^{1}_{\fant{a}2}
&=
0
\\
\end{split}
\qquad \qquad
\begin{split}
\Omega^{0}_{\fant{a}0}
&=
\frac{k_1}{2l_1^2}l\wdg n
\\
\Omega^{3}_{\fant{a}3}
&=
\frac{k_2}{2l_2^2}m\wdg\bar{m}.
\end{split}
\ee
As in  the case of the connection 1-form  the remaining curvature 2-forms are related to  (\ref{product-sym-curvature-forms}) by means of either complex
conjugation or the first Bianchi relations $\eta_{\alpha}\Omega^{\mu}_{\fant{\mu}\beta}+\eta_{\beta\mu}\Omega^{\mu}_{\fant{\mu}\alpha}=0$.
By comparing the result  (\ref{product-sym-curvature-forms}) with the definitions of the curvature spinors as components of curvature 2-forms given in  the Appendix, the above expressions can be  used to obtain the curvature spinors corresponding  to (\ref{background-metrics}) as
\be
\begin{split}
\Phi_{11}
&=
\frac{1}{4}\left(-\frac{k_1}{l_1^2}+\frac{k_2}{l_2^2}\right)
\\
R&=
2\left(\frac{k_1}{l_1^2}+\frac{k_2}{l_2^2}\right)
\end{split}
\qquad
\begin{split}
\Psi_2
&=
-\frac{1}{6}\left(\frac{k_1}{l_1^2}+\frac{k_2}{l_2^2}\right)
\end{split}
\ee
where nonvanishing Weyl spinor $\Psi_2$ indicates that the spaces are not conformally flat and in fact they are of Petrov type D.
For $k_1/l^2_1+k_2/l^2_2\neq0$, the solution to the Einstein field equations require a cosmological term with a constant electromagnetic field.
The Einstein field equations  explicitly takes the form
\be\label{einstein-form-background}
\begin{split}
*G^0&=-\Lambda_+*l+\Lambda_-*l,
\\
*G^1&=-\Lambda_+*n+\Lambda_-*n,
\end{split}
\qquad
\begin{split}
*G^2
&=
-\Lambda_+*m-\Lambda_-*\bar{m},
\\
*G^3
&=
-\Lambda_+*m-\Lambda_-*\bar{m}.
\end{split}
\ee
where the constants $\Lambda_\pm$, which are defined as
\be
\Lambda_+
\equiv
\tfrac{1}{2}(k_1/l^2_1+k_2/l^2_2),
\qquad
\Lambda_-
\equiv
\tfrac{1}{2}(k_1/l^2_1-k_2/l^2_2),
\ee
corresponding to a trace and a trace free part of the Einstein 3-form respectively. Thus, by construction, they
stand for an effective cosmological term and a constant electromagnetic field spinor, respectively (See Eq. (\ref{general-einstein-3forms}) below).
Note that the Ricci curvatures are covariantly constant, $DR^\alpha=dR^\alpha+\omega^{\alpha}_{\fant{a}\beta}\wdg R^\beta=0$ for $\alpha=0, 1, 2, 3$, with  the above curvature and connection expressions of the Ansatz.

In order to write  out the QC field equations for the direct product Ansatz above explicitly, one  first finds the  corresponding auxiliary 2-forms $X^{\alpha\beta}$. In this case,  the explicit expressions for $X^{\alpha\beta}$  reduce to
\be\label{bivector-for-product-space}
\begin{split}
X^{0}_{\fant{a}3}
&=
(4a+b)\Lambda_+l\wdg m
\\
X^{1}_{\fant{a}2}
&=
(4a+b)\Lambda_+n\wdg \bar{m}
\\
\tfrac{1}{2}(X^{0}_{\fant{a}0}-X^{3}_{\fant{a}3})
&=
-
(2a+\tfrac{1}{2}b)\Lambda_+(l\wdg n-m\wdg \bar{m})
-
\tfrac{1}{2}b\Lambda_-(l\wdg n+m\wdg \bar{m})
\end{split}
\ee
because  only the nonvanishing  Ricci spinors are on the diagonal, namely $\Phi_{11}$ and $R$.
Note that relative to a null coframe $\bar{X}^{1}_{\fant{a}2}=X^{1}_{\fant{a}3}$, $\bar{X}^{0}_{\fant{a}3}=X^{0}_{\fant{a}2}$; ${X}^{0}_{\fant{a}0}$ is real while $X^{3}_{\fant{a}3}$ is imaginary. These results imply that,  for (\ref{bivector-for-product-space}), $\Pi^{\alpha\beta}=D*X^{\alpha\beta}=0$ identically for all $\alpha, \beta$. In consequence, the corresponding Lagrange multiplier 2-forms vanish, $\lambda^\alpha=0$ for all $\alpha$. The only contribution of the QC part then comes from $*T^\alpha[X^{\mu\nu}, \Omega^{\mu\nu}]$
\be
\begin{split}
*T^0
&=
-(4a-b)\Lambda_+\Lambda_-*l
\\
*T^1
&=
-(4a-b)\Lambda_+\Lambda_-*n
\end{split}
\qquad
\begin{split}
*T^2
&=
(4a-b)\Lambda_+\Lambda_-
*m
\\
*T^3
&=
(4a-b)\Lambda_+\Lambda_-*\bar{m}.
\end{split}
\ee

The  QC metric equations then admit an electromagnetic field of the form  $\mathcal{F}=\phi_1(l\wdg n-m\wdg \bar{m})$.
For a constant $\phi_1$ and with the background  metric Ansatz (\ref{background-metrics}), the source-free Maxwell's equations are satisfied identically $d\mathcal{F}=0$. Eventually, the metric  field equations for electrovacuum simplify to
 \be
*E^\alpha=-\frac{1}{\kappa^2}*G^\alpha+\frac{1}{\kappa^2}\Lambda_0*\theta^\alpha+*T^\alpha[\Omega^{\mu}_{\fant{a}\nu}, X^{\mu\nu}]+*\tau^\alpha[F]=0
 \ee
and these equations reduce to a system of  equations for the parameters of the metric ansatz (\ref{background-metrics}).
The diagonal components of $E^\alpha$, namely the equations $E^{0}_{\fant{a}0}=E^{1}_{\fant{a}1}=0$ and $E^{2}_{\fant{a}2}=E^{3}_{\fant{a}3}=0$  provide two algebraic equations for the parameters of the ansatz.   For a given $\Lambda_0$ and $\phi_0$ these equations determine the parameters of the Ansatz as
\be\label{backgound-alg-sol}
\begin{split}
\Lambda_+
&=
-\Lambda_0
\\
\Lambda_-
&=
\frac{2\kappa^2\phi_1\bar{\phi}_1}{1+(4a-b)\kappa^2\Lambda_0}.
\end{split}
\ee
Note that  $b=4a$ turns out to be a  peculiar case for the QC part with no apparent physical motivation.
Otherwise, the QC part of (\ref{lag-def})  modifies  the expression for $\Lambda_-$ and that  the equations admit a background solution with $\phi_1\neq0$.
Moreover,  note that pure QC part of the model does not support a cosmological constant because the auxiliary form $*T^\alpha$ has a vanishing trace.

It is worth to emphasize that the background spacetimes  above have  Petrov type D unlike a maximally symmetric solution which  belong to Petrov type O.
Recall that, in four dimensions a constant curvature spacetime with curvature 2-form
\be\label{symmetric-curvature-2form}
\Omega^{\alpha\beta}=-\tfrac{1}{3}\Lambda_0\theta^{\alpha\beta},
\ee
satisfies the Einstein field equations $*G^\alpha=-\Lambda_0*\theta^\alpha$.
These metrics are all conformally flat and thus all Weyl spinors vanish $\Psi_k=0$.
For the  general Lagrangian (\ref{lag-def}),
the maximally symmetric metric   of (\ref{symmetric-curvature-2form}) produces
$*T^{\alpha}=0$.
In this case, the pure QC equations are identically satisfied where the corresponding $D\lambda^\alpha$  and $*T^\alpha$ vanish separately making the constant curvature spaces somewhat trivial solutions to the  QC equations.

\subsection{Impulsive wave solutions in direct product backgrounds}

In this section, Petrov type N impulsive wave solutions will be studied. These waves  propagate in  various type D background spacetimes of the form of a direct product  tabulated in the previous section. The Ansatz for the Maxwell field in the previous section is also valid for the impulsive wave metric Ansatz below.

Shear- and twist-free nondiverging null geodesic vector field and the $pp$-wave Ansatz is a subclass of the Kundt metrics for which a null geodesic vector field is assumed to be  covariantly constant. $pp$-wave space-times admit covariantly constant real null basis 1-form  with vanishing optical scalars and they are space-times are Petrov type N for which all the quadratic curvature invariants namely, $R^2*1, R_\alpha\wdg*R^\alpha, \Omega^{\alpha\beta}\wdg*\Omega_{\alpha\beta}$ vanish.

The wave metrics  that will be discussed  have a form  that are slightly more general then the standard $pp$-wave metric and  they  are introduced on the manifolds which are of the form of  direct product of two two-dimensional manifolds  with an appropriate profile function.  Explicitly, they are assumed to be of  the form
\be\label{metric-ansatz}
g=
\frac{1}{\Omega^2}(du\ot dv+dv\ot du+ 2H(u,\zeta,\bar{\zeta}) du\ot du)
-
\frac{1}{P^2}(d\zeta\ot d\bar{\zeta}+d\bar{\zeta}\ot d{\zeta})
\ee
The real  profile function  is assumed  to be of the distributional-valued  form $H(u,\zeta,\bar{\zeta})=\delta(u)h(\zeta,\bar{\zeta})$ and for $k_1=k_2=0$ the metric (\ref{metric-ansatz}) regain  the Kerr-Schild form of $pp$-wave metric\cite{brinkman,ehlers-kundt}. The metric ansatz is simple enough for the QC field equations yet it slightly more complicated then usual $pp$-wave ansatz that and it leads to a more general equation for the profile function $H$ on a two dimensional manifolds $M_2$. A convenient set of basis  null coframe 1-forms $\{\theta^\alpha\}$ for $\alpha=0,1,2,3$, in conjunction with  the background coframe of the previous section,  consists of the 1-forms
\be\label{pp-coframe}
l=\frac{1}{\Omega}du,\qquad n=\frac{1}{\Omega}(dv+Hdu),\qquad m=\frac{1}{P}d\zeta, \qquad \bar{m}=\frac{1}{P}d\bar{\zeta}.
\ee
The set of frame fields associated to the coframe in (\ref{pp-coframe}) is identical to the set of frame fields \cite{ortaggio} except for the sign of the profile function $H$. As will  be shown below, the metric ansatz (\ref{metric-ansatz}) linearizes the metric field equations  constraining  the constant parameters of the two-spaces $M_1$ and $M_2$.

The connection and the curvature forms corresponding to basis coframe 1-forms (\ref{pp-coframe}) can be calculated as follows. First,
note that in terms of the null coframe 1-forms the exterior derivative operator then takes the form
\be
d
=
l\Omega(\partial_u-H\partial_v)+n\Omega\partial_v+Pm\partial_\zeta+P\bar{m}\partial_{\bar{\zeta}}.
\ee
Using this expression,  the Cartan's first structure equations can be solved for the corresponding connection 1-forms $\omega^{\alpha}_{\fant{a}\beta}$  as
\be
\begin{split}
\omega^{0}_{\fant{a}3}
&=
0
\label{connection-forms}\\
\omega^{1}_{\fant{a}2}
&=
PH_{\zeta} l
\end{split}
\qquad\qquad
\begin{split}
\omega^{0}_{\fant{a}0}
&=
\frac{k_1}{2l_1^2}(vl-un)
\\
\omega^{3}_{\fant{a}3}
&=
\frac{k_2}{2l_2^2}(\bar{\zeta}m-\zeta\bar{m})
\end{split}
\ee
and  by making use of the spin coefficient definitions in the Appendix in (\ref{spinor-def}), the nonvanishing spin coefficients are
\be
\begin{split}
\nu&=-PH_{\zeta}
\\
\epsilon&=\frac{k_2}{2l_1^2}u
\end{split}
\qquad
\qquad
\begin{split}
\gamma
&=-\frac{k_2}{2l_1^2}v
\\
\alpha
&
=
-\bar{\beta}=\frac{k_2}{2l_2^2}\bar{\zeta}
\end{split}
\ee
where coordinate subscript denotes partial derivative with respect to the coordinate.
At this point it is instructive to compare the metric Ansatz (\ref{metric-ansatz}) with the  original $pp$-wave Ansatz. In parallel to the original $pp$-wave metric, the null frame tangent to the real null vector field $l^\sharp=\Delta=e_1$ is twist-free since $l\wdg dl=0$, has a vanishing shear ($\sigma=0$), a vanishing divergence ($\rho=0$)  and it satisfies the geodesic equation
\be
\nabla_{l^\sharp}l=\frac{k_1}{2l_1^2}ul
\ee
since the pseudo-Riemannian covariant derivative $\nabla$ commutes with the linear map $\sharp$ and its inverse. Consequently, the  basis (co)frame field $l$ is not covariantly constant unless $k_1=0.$

The corresponding curvature spinors and the null coframe components of the curvature 2-form can be calculated by using the Cartan's  second structure equations
as
\be\label{curvature-forms}
\begin{split}
\Omega^{0}_{\fant{a}3}
&=
0
\\
\Omega^{1}_{\fant{a}2}
&=
-(P^2H_{\zeta})_{\zeta} l\wdg m -P^2 H_{\zeta\bar{\zeta}}l\wdg \bar{m}
\\
\end{split}
\qquad \qquad
\begin{split}
\Omega^{0}_{\fant{a}0}
&=
-\frac{k_1}{l_1^2}l\wdg n
\\
\Omega^{3}_{\fant{a}3}
&=
-\frac{k_2}{l_2^2}m\wdg\bar{m}
\end{split}
\ee
where the identity $uH=u \delta(u)h(\zeta,\bar{\zeta})=0$ for the distribution-valued  profile function has been used in the derivation.
These curvature expressions reduce to those of $pp$-waves in Minkowski background for $k_1/l^2_1=k_2/l^2_2=1$.
By using the definitions of the  spinor quantities given in  the Appendix, the above curvature expressions can be  used to obtain the corresponding curvature spinors as
\be\label{curvature-spinors}
\begin{split}
\Phi_{11}
&=
\frac{1}{4}\left(-\frac{k_1}{l_1^2}+\frac{k_2}{l_2^2}\right)
\\
\Phi_{22}
&=
\tfrac{1}{2}\Delta H
\end{split}
\qquad
\begin{split}
\Psi_2
&=
-\frac{1}{6}\left(\frac{k_1}{l_1^2}+\frac{k_2}{l_2^2}\right)
\\
\Psi_4
&=
(PH_{\zeta})_{\zeta}
\end{split}
\qquad
R=
2\left(\frac{k_1}{l_1^2}+\frac{k_2}{l_2^2}\right)
\ee
where $\Delta$ is  the two dimensional Laplacian defined on $M_2$ spanned by the   complex coordinates $\{\zeta, \bar{\zeta}\}$ and in terms of local coordinates it  explicitly reads $\Delta\equiv 2P^2\partial_\zeta\partial_{\bar{\zeta}}$.
In comparison to the background spacetimes of the previous section, one has now a nonvanishing Ricci spinor $\Phi_{22}$  and the Weyl spinor $\Psi_4$
which is confined to the null cone defined by $u=0$ representing transverse gravitational waves.
Moreover, these terms do not contribute to $*T^\alpha$. Finally, note that for the $pp$-wave metric the Einstein forms of the background, (\ref{einstein-form-background}) modify accordingly as
\be\label{einstein-form-of-pp-wave}
\begin{split}
*G^0&=-\Lambda_+*l+\Lambda_-*l,
\\
*G^1&=-\Delta H*l-\Lambda_+*n+\Lambda_-*n,
\end{split}
\qquad
\begin{split}
*G^2
&=
-\Lambda_+*m-\Lambda_-*\bar{m},
\\
*G^3
&=
-
\Lambda_+*{m}-\Lambda_-*\bar{m}.
\end{split}
\ee

For $\Lambda_{\mp}=0$, the usual $pp$-wave metric  equation in GR with flat Minkowski background is recovered where the only nonvanishing component of the Ricci tensor $R_{\alpha\beta}$ is that of $l\otimes l$. The above  expressions for the curvature spinors imply that the metric Ansatz lead to constant scalar curvature, unlike the original $pp$-wave metric which has vanishing curvature invariants, and therefore $R=0$.  Note that vanishing scalar curvature renders the introduction of a cosmological constant  incompatible with the  $pp$-wave Ansatz in a Minkowski background.

It is possible to obtain the explicit expressions for the quadratic curvature scalars for the metric Ansatz (\ref{metric-ansatz}). In terms of the
constants in metric ansatz, they are given by
\be
\Omega_{\alpha\beta}\wdg*\Omega^{\alpha\beta}
=
(\Lambda_+^2+\Lambda_-^2)*1
\ee
and
\be
R_{\alpha}\wdg*R^{\alpha}
=
4\Lambda_+\Lambda_-*1
\ee

The geometrical properties of the ansatz given above are sufficient  to calculate
the auxiliary forms for the QC part.  For the $pp$-wave ansatz $X^{\alpha\beta}$ now become
\be\label{bivector-for-pp-space}
\begin{split}
X^{0}_{\fant{a}3}
&=
(4a+b)\Lambda_+l\wdg m
\\
X^{1}_{\fant{a}2}
&=
(4a+b)\Lambda_+n\wdg \bar{m}+\tfrac{1}{2}b\Delta Hl\wdg \bar{m}
\\
\tfrac{1}{2}(X^{0}_{\fant{a}0}-X^{3}_{\fant{a}3})
&=
-
(2a+\tfrac{1}{2}b)\Lambda_+(l\wdg n-m\wdg \bar{m})
-
\tfrac{1}{2}b\Lambda_-(l\wdg n+m\wdg \bar{m}).
\end{split}
\ee
The auxiliary forms of the previous section retain their form, except that $X^{1}_{\fant{a}2}$  receives  an additional $l\wdg m$ component (and hence its complex conjugate $X^{1}_{\fant{a}3}$) in contrast to corresponding auxiliary form $X^{1}_{\fant{a}2}$ of the previous section. Then one can use these auxiliary forms to find an explicit expression for the corresponding Lagrange multiplier form  $\Pi^{\alpha\beta}$. By using the general expressions for the field equations (\ref{general-expression-lag-mult}) for the auxiliary fields,
one finds that $\Pi^{0}_{\fant{a}0}=\Pi^{3}_{\fant{a}3}=\Pi^{0}_{\fant{a}3}=0$ together with  the nonvanishing component
\be\label{Pi-1-2}
\Pi^{1}_{\fant{a}2}=-\tfrac{1}{2}bP(\Delta H)_{,\bar{\zeta}}*l=\bar{\Pi}^{1}_{\fant{a}3}.
\ee

Using the general formula for the covariant exterior derivatives of the vector-valued Lagrange multiplier 2-forms  above, the explicit expressions for the Lagrange multiplier 2-forms can now be calculated using this result.  One finds
\be
\lambda^1
=
-bP[(\Delta H)_{{\zeta}}*(l\wdg m)+(\Delta H)_{\bar{\zeta}}*(l\wdg \bar{m})]
\ee
and $\lambda^0= \lambda^{2}=0$.
Eventually, one arrives at the  expression for the only one nonvanishing fourth-order term
which explicitly reads
\be\label{fourth-order-terms}
D\lambda^1
=
b\left(\Delta\Delta H
+
\frac{k_2}{2l_2^2}P\left[\zeta(\Delta H)_{\zeta}+\bar{\zeta}(\Delta H)_{\bar{\zeta}}\right]\right)*l
\ee
which contributes to the  $l\ot l$ component of $E_{\alpha\beta}$ (corresponding to  the off-diagonal component $E^{1}_{\fant{a}0}$).
The equations $*E^\alpha=0$  (\ref{general-metric-eqns}) naturally separate the fourth-order Lagrange multiplier part $D\lambda^\alpha$ that are linear in curvatures  and the part that are nonlinear in curvature and the background quantities are involved only in the latter part. The background solutions, which are studied in the previous section, are remained untouched  and do  not involve in the expression for $\lambda^\alpha$. Obviously, for the impulsive wave Ansatz above, the pure QC part of the model gives rise to a term that is linear in the profile function $H$ appearing in the off-diagonal term (\ref{fourth-order-terms}). The last two terms  on the right-hand side of (\ref{fourth-order-terms}) vanish for $k_2=0$ and accordingly with $\Delta\mapsto 2\partial_\zeta\partial_{\bar{\zeta}}$, the equations for the $pp$ wave  with flat background are obtained.

Using the expression for $G^{1}_{\fant{a}0}$ in (\ref{einstein-form-of-pp-wave}) and assuming that the parameters of the background configuration are satisfied, the only equations for the impulsive wave metric Ansatz (\ref{metric-ansatz}) reduces to $E^{1}_{\fant{a}0}=0$ and this equation explicitly reads
\be\label{general-profile-eqn}
\Delta
\left(\Delta +L^{-2}\right)h
+
(\Lambda_+-\Lambda_-)P[\zeta(\Delta h)_{,\zeta}+\bar{\zeta}(\Delta h)_{,\bar{\zeta}}]
=
0
\ee
with the constant $L^{-2}\equiv(b\kappa^{2})^{-1}$. Apparently, a slight modification of $pp$-metric ansatz entails a more complicated partial differential equation  for the corresponding profile function.

Provided that the profile function, which is assumed to be of the form  $H(u, \zeta,\bar{\zeta})=\delta(u)h(\zeta, \bar{\zeta})$,  satisfies the complex
fourth-order partial differential equation (\ref{general-profile-eqn}), the metric (\ref{metric-ansatz}) defines a family of impulsive gravitational wave solution propagating in Petrov type D backgrounds to the gravitational model based on the Lagrangian (\ref{lag-def}). The second term in (\ref{general-profile-eqn}) is related to the type D background and it  drops out for $k_2=0$ in accordance with the previous result,  cf., for example, \cite{buchdahl}. Note that the  $M_1$ part of the metric ansatz and, hence the metric function $\Omega$, does not enter into the equation for the profile function (\ref{general-profile-eqn}) and show up only in the equations for the background therefore. The second term in (\ref{general-profile-eqn})  does not appear on the  Plebanski-Hacyan background spacetime of the form $AdS_2\times \mathbb{R}^2$ whereas  for Nariai (anti-Nariai) type background, (\ref{general-profile-eqn}) is an equation on $S_2 (H_2)$. It is worth to emphasize that at this point that the set of solutions to the equation (\ref{general-profile-eqn}) defines a new family of algebraically special  solutions to the QC model based on the total Lagrangian (\ref{lag-def}).

The  familiar vacuum $pp$-waves equations for the vacuum  on flat background is recovered by setting $k_1=k_2=0$, $\Lambda_0=0$ and $F=0$. In this  subcase, $\alpha=1$ component of the metric field equations  (\ref{general-metric-eqns}) reduce  to
\be\label{pp-wave-qc-flat-background}
*E^1
=
\left(bd*d+\kappa^{-2}*\right)R^1=0.
\ee
The expression  on the right side in (\ref{pp-wave-qc-flat-background})  follows  from $*T^1[\Omega_{\mu\nu}, X^{\mu\nu}]=0$, $DR^1=dR^1$ and $D\lambda^1=d\lambda^1$ for the $pp$-metric with flat background with $R^1=-2H_{\zeta\bar{\zeta}}l$. The field equation (\ref{pp-wave-qc-flat-background}) reflects the efficiency of the coframe formulation and the exterior form language, in that it is possible to show by construction that the original $pp$-wave metric linearize the QC field equations in a straightforward manner by retaining the terms linear in curvature, namely the term $D*DR^\alpha$ in the simplified form $d*dR^\alpha$ since the only nonzero connection 1-form is $\omega^{1}_{\fant{a}2}$ having  a single nonzero component (see also the discussion in Section C of the Appendix). In addition, it is straightforward to show   that for the $pp$-wave metric the operator $d*d$ on $M_1\times M_2\cong \mathbb{R}^{(1,3)}$ becomes the Laplacian operator on $M_2\cong \mathbb{R}^2$ acting on the profile function $H(u,\zeta, \bar{\zeta})$.

\section{Concluding remarks}

The current work introduces  a new family of impulsive gravitational waves propagating in various product background spacetimes for  a general QC gravity model based on the Lagrangian form (\ref{lag-def}) in four dimensions. The gravitational sector of the model   contains an Einstein-Hilbert term, a cosmological  constant, and the quadratic curvature terms. The only matter coupling to the geometry is assumed to be the electromagnetic field with the Lagrangian $-\tfrac{1}{2}F\wdg *F=-\tfrac{1}{4}F_{\alpha\beta}F^{\alpha\beta}*1=-\tfrac{1}{4}F_{ab}F^{ab}*1$.
With the help of the expression for the inner product $R_\alpha\wdg *R^\alpha=R_{\alpha\beta}R^{\alpha}_{\fant{a}\mu}\theta^\beta\wdg *\theta^\mu=R_{\alpha\beta}R^{\alpha\beta}*1$ which follows from $\theta^\beta\wdg *\theta^\mu=\eta^{\beta\mu}*1$, the Lagrangian density (\ref{lag-def}) can be rewritten in the form
\be\label{lag-def2}
\mathcal{L}
=
\left(\frac{1}{2\kappa^2}R
+
\frac{1}{\kappa^2}\Lambda_0
+
\frac{a}{2}R^2+\frac{b}{2} R_{ab}R^{ab}-\frac{1}{4}F_{ab}F^{ab}\right)\sqrt{|g|}dx^0\wdg dx^1\wdg dx^2\wdg dx^3
\ee
where $R_{ab}$ stands for the components of the  Ricci tensor relative to a coordinate frame,  $R=\eta^{\alpha\beta}R_{\alpha\beta}=g^{ab}R_{ab}$ is the scalar curvature  and the invariant volume element is written in the form $*1=\sqrt{|g|}dx^0\wdg dx^1\wdg dx^2\wdg dx^3$.

The field equations that follow from  the Lagrangian form (\ref{lag-def2})  admit solutions with the  metric of the form
\be\label{metric-ansatz2}
g
=
\frac{du\ot dv+dv\ot du+ 2\delta(u)h(\zeta, \bar{\zeta})du\ot du}{(1-\frac{k_1}{l_1^2}uv)^2}
-
\frac{d\zeta\ot d\bar{\zeta}+d\bar{\zeta}\ot d\zeta}{(1+\frac{k_2}{l_2^2}\zeta\bar{\zeta})^2}
\ee
which is  slightly more general then the familiar  Kerr-Schild form of the impulsive $pp$-wave metric.
For a vanishing profile function, the background spacetimes   of the product form $M_1\times M_2$ are solutions to the field equations
if the constants  $k_1=l_1^{2}(\Lambda_++\Lambda_-)$ and $k_2=l_2^{2}(\Lambda_+-\Lambda_-)$ are determined by the relations
\begin{align}
\Lambda_+
&=
-\Lambda_0
\label{Lambda+}\\
\Lambda_-
&=
\frac{2\kappa^2\phi_1\bar{\phi}_1}{1+(4a-b)\kappa^2\Lambda_0}\label{Lambda-}
\end{align}
in terms the cosmological constant $\Lambda_0$,  the constant electromagnetic field spinors $\phi_1$ and the QC coupling constants $a, b$.
These background spacetimes provide a family of algebraically special solutions to the QC  model defined by (\ref{lag-def2}).
Furthermore, the distributional-valued profile function $\delta(u)h(\zeta,\bar{\zeta})$ appears only in the terms that are linear in the derivatives of the curvature components in the field equations, and thus $h$ satisfies the following fourth-order equation on $M_2$
\be
\Delta
\left(\Delta +L^{-2}\right)h
+
(\Lambda_+-\Lambda_-)P[\zeta(\Delta h)_{,\zeta}+\bar{\zeta}(\Delta h)_{,\bar{\zeta}}]
=
0
\ee
with $L^2\equiv b\kappa^2$, $\Delta\equiv2P^2\partial_\zeta\partial_{\bar{\zeta}}$ and $\Lambda_{\mp}$ are given as in
(\ref{Lambda+}) and (\ref{Lambda-}). Consequently, the impulsive wave solutions  defined by the metric  (\ref{metric-ansatz2})  for various  product backgrounds  has the property that the resulting field equation is linear in the profile function with the parameters of the two dimensional manifolds satisfying  some algebraic equations.

The evaluation of the field equations relative to a suitable null coframe   by using an impulsive wave-type metric Ansatz   presented above provides  an example for the scheme of calculations developed and it is possible to write out the metric field equations $*E^\alpha=0$ in terms  the spinor  quantities by using Eqs. (\ref{curv-spin-defs})  and their appropriate contractions in full generality. Thus, the formulas of Sec. II provides a practical and alternative approach  for  more complicated metric Ansatz towards the efforts in finding the solutions to the  QC field equation by taking algebraic type into the account.

The following remarks regarding  the breadth of the applicability of the null coframe formalism developed above can be stated.

Although the discussion  in this work is confined to four spacetime dimensions, the   mathematical framework of the NP spin coefficient formalism expressed in the language  of exterior differential forms has the prospect in generalization to higher dimensions which needs further scrutiny of the above formulas in conjunction with the previous work \cite{pravda-pravdova}. In this regard, the use of exterior algebra  may provide formal simplification for  the extension of Newman-Penrose formalism to higher dimensions recently studied in a series of papers \cite{pravda-pravdova-coley-milson,pravda-pravdova,coley-milson-pravda-pravdova,ortaggio-pravda-pravdova}.

It is possible to extend the above discussion to three  spacetime dimensions as well. In particular, the new massive gravity Lagrangian \cite{nmg} with cosmological constant can be obtaining by setting $a=-\frac{3}{8}b$ in the Lagrangian  (\ref{lag-def}).   The field equations  for new massive gravity expressed in terms of Cotton 2-form \cite{baykal} then can be formulated in terms of a seminull coframe by making use of spinor formulation of topologically massive gravity introduced  in \cite{aliev-nutku}.

\section*{Appendix}

The following set of formulas provides a glossary of the NP quantities  in relation to the associated tensorial  quantities used in the main text, see, for example, \cite{exact-sol-stephani} for the discussion in terms of a set  of null frame fields. The spinor definitions in terms of the components of the tensor-valued forms and the exterior forms are in accordance  with the original NP definitions \cite{newman-penrose}. The numerical indices refer  to a complex null coframe throughout the Appendix.

\subsection{NP spinor definitions}

The notation used in the paper allows one to carry out  tensorial calculations using the algebra  of exteriors forms
relative to a local null coframe  denoted by $\{\theta^\alpha\}=\{l, n, m, \bar{m}\}$ for $\alpha=0, 1, 2, 3$ respectively.
$l,n$ is a pair of real null vectors whereas $m, \bar{m}$ is a pair of complex-conjugate spatial null vectors.
(A bar over a quantity denotes complex conjugation).  A set of null coframe 1-forms $\{\theta^\alpha\}$ defines a set of associated frame vectors
$\{e_\alpha\}=\{n^\sharp, l^\sharp, -\bar{m}^\sharp, -m^\sharp\}$ which are denoted by $\{\Delta, D, -\bar{\delta}, -\delta\}$ in the NP spin coefficient formalism, respectively. The isomorphism $\sharp$ maps a coframe 1-form to an associated basis frame field.
$n^\sharp$ is the vector field associated with the 1-form $n=n_adx^a$, i.e., $n^\sharp=n^a\partial_a$ with $n^{a}=g^{ab}n_b$.
Relative to a null frame, the contraction operator can be written conveniently using the map $\sharp$, for example,
$i_{e_0}\equiv i_0=i_{n^\sharp}$. The  definition of the map $\sharp$ facilitates the calculation of the contractions, such as
$i_{n^\sharp} l=l_an^a=g_{ab}n^al^b=1$
where  $g=g_{ab}dx^a\otimes dx^b$.

The numerical indices relative to the null coframe are raised and lowered by the metric with nonvanishing components $\eta_{01}=-\eta_{23}=1$ and
$\eta^{01}=-\eta^{23}=1$. For example, $G_3=\eta_{3\alpha}G^{\alpha}=-G^2$  and as another example, for the indices of the curvature 2-form,
one has  $\Omega^{0}_{\fant{a}0}=\Omega_{10}=\Omega^{01}$, etc..

The Hodge dual operator acting on a $p$-form is denoted by $*$. In terms of  a NP null coframe, the orientation is chosen so that the invariant volume 4-form takes the form $*1=-il\wdg n\wdg m\wdg\bar{m}$. Accordingly, the corresponding permutation symbol can be defined as  $l^an^bm^c\bar{m}^d\epsilon_{abcd}=\epsilon_{0123}=-i$ can also be employed to calculate Hodge duals relative to null coframe. The Hodge duality relations for the basis $p$-forms then follows from the relations
\be
\begin{split}
*l
&=
+il\wdg m\wdg \bar{m}
\\
*n
&=
-in\wdg m\wdg \bar{m}
\\
*m
&=
-il\wdg n\wdg {m}
\end{split}
\qquad\qquad
\begin{split}
*(l\wdg n)
&=
+im\wdg \bar{m}
\\
*(l\wdg m)
&=
-il\wdg m
\\
*(n\wdg \bar{m})
&=
-in\wdg \bar{m}
\end{split}
\ee
where the Hodge duals of the basis 3-forms can be found by using $**=id$ acting on 3-forms and in general for any $p$-form
$**=(-1)^{p(n-p)+s}id$, where $s$ is signature of the metric.

In connection with  the Hodge dual, the contraction operator $i_X$,  and the wedge product $\wdg$,
the identities $i_X*\omega=*(\omega\wdg X^\flat)$ and $X^\flat\wdg *\omega=(-1)^{(p+1)}*i_X\omega $ are frequently used in tensorial manipulations as well as in explicit computations. Here $X$ is a vector field, $\omega$ is a $p$-form, and $\flat$ is the inverse of the map $\sharp$ \cite{straumann}.

In the present mathematical framework, the  expressions for the NP spinor scalars  \cite{newman-penrose} can be  read off from an associated
tensorial expressions. In particular, the  twelve complex NP spin coefficients correspond to  the following components of the Levi-Civita connection 1-form $\omega^{\alpha}_{\fant{a}\beta}$:
\be\label{spinor-def}
\begin{split}
\omega^{0}_{\fant{0}3}
&=
+\tau l+\kappa n-\rho  m-\sigma \bar{m}
\\
{\omega}^{1}_{\fant{0}2}
&=
-\nu l-\pi n+\lambda m+\mu\bar{m}
\\
\tfrac{1}{2}(\omega^{0}_{\fant{0}0}-\omega^{3}_{\fant{0}3})
&=
-\gamma l-\epsilon n+\alpha m+\beta \bar{m}.
\end{split}
\ee
Relative to an orthonormal coframe, there are six independent connection 1-forms, whereas relative to a NP null coframe there are
only three complex connections. In either case, and for an arbitrary vector field $X$, they can also be defined by $\nabla_X\theta^\alpha=-i_X(\omega^{\alpha}_{\fant{a}\beta})\theta^\beta$ from which the original definitions of the spin coefficient expressions   relative to a coordinate basis in terms of covariant derivatives can be recovered. The covariant exterior derivative denoted by $D$ can be derived from the definition of $\nabla_X$, and acting on tensor-valued forms,  it is often more convenient to use.

The spin coefficient  definitions (\ref{spinor-def}) can be used in Cartan's first structure equations to find the exterior derivatives of the basis coframe 1-forms as \cite{penrose-rindler}
\be\label{exterior-derivative-cof}
\begin{split}
&dl
+
(\epsilon+\bar{\epsilon})l\wdg n-(\alpha+\bar{\beta}-\bar{\tau})l\wdg m-(\bar{\alpha}+\beta-{\tau})l\wdg \bar{m}
+
\bar{\kappa} n\wdg m
+
{\kappa} n\wdg \bar{m}
-
(\rho-\bar{\rho})m\wdg \bar{m}
=0
\\
&
dn
+
(\gamma+\bar{\gamma})l\wdg n
+
(\alpha+\bar{\beta}-\pi)n\wdg m
+
(\bar{\alpha}+{\beta}-\bar{\pi})n\wdg \bar{m}
-
\nu l\wdg m
-
\bar{\nu}l\wdg \bar{m}
-
(\mu-\bar{\mu})m\wdg\bar{m}
=0
\\
&dm
+
(\bar{\pi}+{\tau})l\wdg n
-
(\gamma-\bar{\gamma}+\bar{\mu})l\wdg m
-
\bar{\lambda}
l\wdg \bar{m}
-
(\epsilon-\bar{\epsilon}-\rho)
n\wdg m
+
\sigma
n\wdg \bar{m}
+
(\bar{\alpha}-\beta)
m\wdg \bar{m}
=0.
\end{split}
\ee

These 2-form equations  are completely equivalent in content to the commutator relations $[\Delta, D], [\Delta, \delta], [D, \delta], [\delta, \bar{\delta}]$ for the basis vector fields.  The commutators can be used to derive the commutators by making use of the operator identity $[L_X,i_Y]=i_{[X,Y]}$ acting on  null basis 1-forms for the Lie derivative $L_X=di_X+i_Xd$ and the contraction operator \cite{straumann}.

In terms of a  null coframe, the exterior derivative operator, denoted by  $d$, has the general expression
\be\label{ext-der-gen}
d
=
l\Delta+nD-m\bar{\delta}-\bar{m}\delta
\ee
acting on scalars. The expressions (\ref{exterior-derivative-cof}) and (\ref{ext-der-gen})
can be used to   calculate the exterior derivative of an arbitrary $p$-form relative to a given null coframe.
Note that (\ref{exterior-derivative-cof})  provides an algebraic system of equations for calculation of the spin coefficients.

As a more practical alternative, it is possible to obtain a general formulas for the NP spin coefficients by making use of
Cartan's first structure equation
\be
\Theta^\alpha
=
D\theta^\alpha
=
0
=
d\theta^\alpha+\omega^{\alpha}_{\fant{a}\beta}\wdg \theta^\beta
\ee
where $\Theta^\alpha=\frac{1}{2}T^{\alpha}_{\fant{a}\mu\nu}\theta^{\mu\nu}$ is torsion 2-form with  $T^{\alpha}_{\fant{a}\mu\nu}$ denoting
the components of the torsion tensor. With the assumptions of the vanishing  torsion 2-form and nonmetricity,
these equations can be solved for the Levi-Civita connection 1-forms \cite{thirring} to find
\be
\omega^{\alpha}_{\fant{a}\beta}
=
\tfrac{1}{2}i^\alpha i_\beta(d\theta^\mu\wdg \theta_\mu)
-
i^\alpha d\theta_\beta+i_\beta d\theta^\alpha.
\ee
For a  concise derivation of this formula using the exterior algebra, see, for example, \cite{baykal-epjp}.
The general formula is valid in a null or orthonormal coframe and in particular, relative to a null coframe and with due attention paid to the signs, one can find the following expressions:
\be\label{connection-1-forms-extra}
\begin{split}
\omega^{0}_{\fant{q}3}
&=
-\tfrac{1}{2}i_{l^\sharp}i_{m^\sharp}(l\wdg dn+n\wdg dl-m\wdg d\bar{m}-\bar{m}\wdg dm)+i_{l^\sharp}dm-i_{m^\sharp}dl
\\
\omega^{1}_{\fant{q}2}
&=
-\tfrac{1}{2}i_{n^\sharp}i_{\bar{m}^\sharp}(l\wdg dn+n\wdg dl-m\wdg d\bar{m}-\bar{m}\wdg dm)+i_{n^\sharp}d\bar{m}-i_{\bar{m}^\sharp}dn
\\
\omega^{0}_{\fant{q}0}
&=
+\tfrac{1}{2}i_{l^\sharp}i_{n^\sharp}(l\wdg dn+n\wdg dl-m\wdg d\bar{m}-\bar{m}\wdg dm)-i_{l^\sharp}dn+i_{n^\sharp}dl
\\
\omega^{3}_{\fant{q}3}
&=
-\tfrac{1}{2}i_{\bar{m}^\sharp}i_{m^\sharp}(l\wdg dn+n\wdg dl-m\wdg d\bar{m}-\bar{m}\wdg dm)+i_{\bar{m}^\sharp}dm-i_{m^\sharp}d\bar{m}.
\end{split}
\ee
for the connection 1-forms used in the discussion above.
The general formulas in (\ref{connection-1-forms-extra}) reduce the main labor in calculation of the spin coefficients (i.e., the Ricci rotation coefficients) to the calculation of the exterior derivatives $dl, dn$, and $dm$ with the help of (\ref{ext-der-gen}) and expressing in terms of the basis 2-forms, namely $l\wdg n$, $l\wdg m$, $l\wdg \bar{m}$, $n\wdg m$, $n\wdg \bar{m}$  and $m\wdg \bar{m}$. Subsequently, by carrying out the contractions indicated in (\ref{connection-1-forms-extra}) and then identifying the components of the resultant  expression with (\ref{spinor-def}), it is straightforward to find the  NP spin coefficients by hand.

The conventions for the curvature spinors also follow the original NP spin coefficient formalism  to facilitate the comparison with the literature \cite{newman-penrose}. In particular, the Ricci 1-form can conveniently be defined by  $R^\alpha\equiv i_\beta\Omega^{\alpha\beta}$ in terms of the contraction of the curvature 2-form $\Omega^{\alpha}_{\fant{a}\beta}=\frac{1}{2}R^{\alpha}_{\fant{a}\beta\mu\nu}\theta^{\mu\nu}$ and that the Ricci 1-form explicitly reads $R^{\alpha}=R^{\alpha\mu}_{\fant{aa}\mu\beta}\theta^\beta$. The scalar curvature can be defined as the contraction $R=i_\alpha R^\alpha$. In terms of the curvature 2-forms, Cartan's  second structure equation reads
\be\label{cartan-SE2}
\Omega^{\alpha}_{\fant{a}\beta}
=
d\omega^{\alpha}_{\fant{a}\beta}
+
\omega^{\alpha}_{\fant{a}\mu}
\wdg
\omega^{\mu}_{\fant{a}\beta}.
\ee
In terms of the connection forms above, (\ref{cartan-SE2}) explicitly reads
\be\label{cartan-SE2-form2}
\begin{split}
{\Omega}^{0}_{\fant{0}3}
&=
d{\omega}^{0}_{\fant{0}3}
-
{\omega}^{0}_{\fant{0}3}\wedge\left(\omega^{0}_{\fant{0}0}
-
\omega^{3}_{\fant{0}3}\right)
\\
{\Omega}^{1}_{\fant{0}2}
&=
d{\omega}^{1}_{\fant{0}2}
+
{\omega}^{1}_{\fant{0}2}\wdg\left(\omega^{0}_{\fant{0}0}
-
\omega^{3}_{\fant{0}3}\right)
\\
\Omega^{0}_{\fant{0}0}-\Omega^{3}_{\fant{0}3}
&=
d(\omega^{0}_{\fant{0}0}-\omega^{3}_{\fant{0}3})
+
2{\omega}^{0}_{\fant{0}3}\wedge{\omega}^{1}_{\fant{0}2}.
\end{split}
\ee
The 18 scalar field equations, namely (4.2a)-(4.2r) of \cite{newman-penrose},  involving the curvature spinors (Ricci spinors $\Phi_{ik}$, Weyl spinors $\Psi_k$  and scalar curvature $R$) can be derived from  Cartan's second structure equation (\ref{cartan-SE2-form2}), by using  (\ref{spinor-def}) together with the following definition of curvature spinors
\be
\begin{split}
{\Omega}^{0}_{\fant{0}3}
&=
C^{0}_{\fant{0}3}
+
[-\Phi_{00}n\wdg m-\Phi_{01}(l\wdg  n+m\wdg \bar{m})+\Phi_{02}l\wdg \bar{m}]
+
\tfrac{1}{12} Rl\wdg m
\label{curv-spin-defs}\\
{\Omega}^{1}_{\fant{0}2}
&=
C^{1}_{\fant{0}2}
+
[
\Phi_{20}n\wdg m
+
\Phi_{21}(l\wdg n+m\wdg\bar{m})
-
\Phi_{22}l\wdg \bar{m}
]
+
\tfrac{1}{12}Rn\wdg \bar{m}
\\
\tfrac{1}{2}({\Omega}^{0}_{\fant{0}0}-{\Omega}^{3}_{\fant{0}3})
&=
\tfrac{1}{2}(C^{0}_{\fant{0}0}-C^{3}_{\fant{0}3})
+
[
\Phi_{10}n\wdg m
+
\Phi_{11}(l\wdg n+m\wdg\bar{m})
-
\Phi_{12}l\wdg \bar{m}
]
-
\tfrac{1}{24}R(l\wdg n-m\wdg\bar{m})
\end{split}
\ee
where $C^{\alpha}_{\fant{a}\beta}$ are the Weyl curvature 2-forms with the following identification of components
\be\label{weyl-spinor-def}
\begin{split}
C^{0}_{\fant{z}3}
&=
-
\Psi_{0}n\wdg \bar{m}
-
\Psi_{1}(l\wdg n-m\wdg \bar{m})
+
\Psi_{2}l\wdg m
\\
C^{1}_{\fant{z}2}
&=
+
\Psi_{2}n\wdg \bar{m}
+
\Psi_{3}(l\wdg n-m\wdg \bar{m})
-
\Psi_{4}l\wdg m
\\
\tfrac{1}{2}(C^{0}_{\fant{z}0}-C^{3}_{\fant{z}3})
&=
+
\Psi_{1}n\wdg \bar{m}
+
\Psi_{2}(l\wdg n-m\wdg \bar{m})
-
\Psi_{3}l\wdg m
\end{split}
\ee
in terms of Weyl spinors $\Psi_k$.
The curvature spinor definitions (\ref{curv-spin-defs}) and (\ref{weyl-spinor-def}) follow from the   familiar decomposition of the
curvature 2-form into the parts that are  irreducible representations of the Lorentz group as
\be\label{curvature-decomp}
{\Omega}^{\alpha}_{\fant{0}\beta}
=
C^{\alpha}_{\fant{0}\beta}
-
\tfrac{1}{2}(\theta^{\alpha}\wdg S_{\beta}-\theta^{\beta}\wdg S_{\alpha})
-
\tfrac{1}{12}R\theta^{\alpha}\wdg \theta_{\beta}
\ee
with the indices specialized relative to an NP coframe.  The Weyl 2-form, denoted by $C^{\alpha}_{\fant{0}\beta}$, is an irreducible traceless fourth rank part, whereas  $S^\alpha=R^\alpha-\frac{1}{4}R\theta^\alpha$ is the second rank traceless Ricci 1-form part, and the remaining term  $R$ is the trace part.
The second rank part in anti-self-dual whereas the remaining parts  comprise the self-dual parts. For an elegant discussion of the selfdual and the anti-selfduality properties of the irreducible parts in terms of differential forms, see for example \cite{obukhov-curvature-decomp}.

There are 11 complex  scalar equations for the second Bianchi identity in the NP formalism and, in general, in terms of the covariant exterior derivative of the curvature 2-form it reads $D\Omega^{\alpha}_{\fant{a}\beta}=0$. More precisely, the components of the 3-form equations $D\Omega^{0}_{\fant{a}3}=0$ (four scalar equations), $D\Omega^{1}_{\fant{a}2}=0$ (four scalar equations), and $D*G^\alpha=0$ (two real scalar equations for $\alpha=0, 1$ and a complex scalar equation for $\alpha=2$) comprise the Bianchi identity expressed as scalar equations in the NP formalism. The formulas in terms of exterior forms  to this point then embody the field equations in the NP spin coefficient formalism succinctly.

The Einstein field equations, $*G^\alpha=\kappa^2*\tau^\alpha$, expressed in the above notation,  allow one to replace Ricci  spinors with the components of  matter energy-momentum forms $*\tau^\alpha$ relative to null coframe ($\kappa^2\equiv 8\pi Gc^{-4}$). For a given matter energy-momentum content, the Ricci spinor components can be  obtained from the Einstein field equations, provided that one has the following identification of the components of the Einstein  3-forms in terms of Ricci spinors relative to a NP null coframe,
\be\label{general-einstein-3forms}
\begin{split}
*G^0
&=
+
i\bar{\Omega}^{0}_{\fant{1}3}\wdg m
-
i{\Omega}^{0}_{\fant{1}3}\wdg \bar{m}
+
i\Omega^{3}_{\fant{1}3}\wdg l
\\
&=
-
2(\Phi_{11}+\tfrac{1}{8}R)*l
-
2\Phi_{00}*n
+
2\Phi_{10}*m
+
2\Phi_{01}*\bar{m}
\\
*G^1
&=
-
i\Omega^{1}_{\fant{1}2}\wdg m
+
i\bar{\Omega}^{1}_{\fant{1}2}\wdg \bar{m}
-
i\Omega^{3}_{\fant{1}3}\wdg n
\\
&=
-
2\Phi_{22}*l
-
2(\Phi_{11}+\tfrac{1}{8}R)*n
+
2\Phi_{21}*m
+
2\Phi_{12}*\bar{m}
\\
*G^2
&=
-
i\Omega^{0}_{\fant{1}0}\wdg m
+
i\bar{\Omega}^{1}_{\fant{1}2}\wdg l
-
i\Omega^{0}_{\fant{1}3}\wdg n
\\
&=
-
2\Phi_{12}*l
-
2\Phi_{01}*n
+
2(\Phi_{11}-\tfrac{1}{8}R)*m
+
2\Phi_{02}*\bar{m}
\\
*G^3
&=
i\Omega^{0}_{\fant{1}0}\wdg \bar{m}
-
i\Omega^{1}_{\fant{1}2}\wdg l
+
i\bar{\Omega}^{0}_{\fant{1}3}\wdg n
\\
&=
-
2\Phi_{21}*l
-
2\Phi_{10}*n
+
2\Phi_{20}*m
+
2(\Phi_{11}-\tfrac{1}{8}R)*\bar{m}
\end{split}
\ee
where $G^\alpha\equiv G^{\alpha}_{\fant{a}\beta}\theta^\beta=(R^{\alpha}_{\fant{a}\beta}-\tfrac{1}{2}\delta^{\alpha}_{\beta}R)\theta^\beta$.  In actual calculations,  explicit expressions for the Einstein 3-forms  can  be derived  from the expressions on the right-hand side in (\ref{general-einstein-3forms}) in terms of the curvature 2-forms. At the same time, (\ref{general-einstein-3forms}) can also be used to identify the tensorial components  of the traceless Ricci 1-forms in terms of  Ricci spinors $\Phi_{ik}$. Even a more compact null coframe formulation of the NP field  equations and the Einstein 3-forms was introduced in \cite{gurses,guven} by utilizing matrix-valued differential forms. Yet another  null coframe approach was introduced in \cite{dereli-tucker-PLA} by making use of complex quaternionic differential forms.

\subsection{Maxwell spinor definitions}

Because the NP formalism makes  use of complex geometrical quantities defined relative to a null coframe, any matter source should be projected to the complex null coframe accordingly. Relative to a null coframe,  Maxwell's  equations can be written by introducing the complex 2-form
\be\label{self-dual-F}
\mathcal{F}
=
\tfrac{1}{2}(F+i*F)
\ee
which is a self-dual 3-form  $*\mathcal{F}=-i\mathcal{F}$ by definition. In terms of $\mathcal{F}$, the original Maxwell spinor  definitions of NP
follow  from
\be
\mathcal{F}
=
\phi_0 n\wdg \bar{m}+\phi_1(l\wdg n-m\wdg \bar{m})-\phi_2 l\wdg m
\ee
where the  basis 2-forms $n\wdg \bar{m}$, $l\wdg m$, and $l\wdg n-m\wdg \bar{m}$ are self-dual whereas their complex conjugates are anti-self-dual.
NP form of the Maxwell's equations, the set of scalar equations (A1) in \cite{newman-penrose}, in terms of complex spinors $\phi_k$ and the spin coefficients, follow from the equation $d\mathcal{F}=0$ vanishing componentwise.
Finally, with the help of the definition (\ref{self-dual-F}), the energy-momentum form of the Maxwell field (\ref{en-mom-F}) can be written as
\begin{align}
*\tau_\alpha[F]
&=
\tfrac{1}{2}[(i_\alpha F)\wdg *F-F\wdg i_\alpha *F]
\nonumber\\
&=
2i(i_\alpha \mathcal{F})\wdg\bar{\mathcal{F}}\label{en-mom-F-2-form}
\end{align}
where the expression in the second line is sometimes more convenient and  $\tau_{\alpha\beta}=2\mathcal{F}_{\alpha\mu}\mathcal{F}^{\mu}_{\fant{a}\beta}$. By combining the definitions (\ref{en-mom-F-2-form}) and (\ref{general-einstein-3forms}), one can show that the Einstein-Maxwell  equations $*G^\alpha=\kappa^2*\tau^\alpha[F]$ then determine the Ricci spinors as $\Phi_{ik}=\kappa^2\phi_i\bar{\phi}_k$. Subsequently, these relations are to be inserted into the structure equations  (\ref{curv-spin-defs}) or into the scalar NP field equations (commonly referred to as Ricci identities) which constitute the components of the tensorial relation given in (\ref{curv-spin-defs}).  Therefore, as in the Einstein-Maxwell case, the null coframe formalism in general allows one to consider the components of the Weyl 2-form together  with the Ricci tensor components  at the same time at the level of the structure equations. This is, in fact, one of the reasons why NP formalism provides a mathematically robust and convenient mathematical framework to discuss algebraically special metrics in contrast to the usual tensorial methods.

\subsection{Prime symmetry}

The prime symmetry provides valuable and practical consistency checks for the field equations in the  NP formalism.
As a discrete symmetry, it can be defined as the exchange symmetry $l\leftrightarrow n$ and $m\leftrightarrow\bar{m}$ for a basis coframe 1-forms. It
can be considered as a map  of the complex Cartan's structure equations onto themselves and therefore can be extended to the tensorial objects and their components in the exterior algebra. In  terms of the numerical indices belonging to a null coframe, it amounts to the exchanges $0\leftrightarrow 1$ and $2\leftrightarrow3$. For example, for the curvature 2-forms, it corresponds  to the exchanges $\Omega^{0}_{\fant{a}3}\leftrightarrow\Omega^{1}_{\fant{a}2}$, $\Omega^{0}_{\fant{a}0}\leftrightarrow \Omega^{1}_{\fant{a}1}$  and $\Omega^{2}_{\fant{a}2}\leftrightarrow\Omega^{3}_{\fant{a}3}$. The prime symmetries of the spin coefficients and the curvature spinors follow from the associated tensorial prime symmetry.

For a $pp$-wave metric Ansatz (\ref{metric-ansatz}) with $\Omega=P=1$, the associated  coframe can be chosen in two different ways that are related by the prime     symmetry. In this case, the prime symmetric  companion of the coframe (\ref{pp-coframe}) leads to  $\omega^{1}_{\fant{a}2}\mapsto\omega^{0}_{\fant{3}3}=H_{\bar{\zeta}}n$, $\omega^{0}_{\fant{a}3}\mapsto\omega^{1}_{\fant{3}2}=0$. Consequently,  the nonvanishing prime companion curvature 2-forms are  $\Omega^{1}_{\fant{a}2}\mapsto \Omega^{0}_{\fant{a}3}=-H_{{\zeta}\bar{\zeta}} n\wdg m -H_{\bar{\zeta}\bar{\zeta}}n\wdg\bar{m}$. Accordingly, the corresponding Einstein form  $*G^1=*R^1=-2H_{\zeta\bar{\zeta}}*l=-d*dn$  is prime companion to $*G^0=*R^0-2H_{\zeta\bar{\zeta}}*n=-d*dl$ as well.

The prime symmetry relating  for the two possible choices of the basis coframes for the  $pp$-wave metric can be extended to the auxiliary tensor-valued forms defined above for the QC equations. For example, one has the prime symmetry companions $\Pi^{1}_{\fant{a}2}\leftrightarrow\Pi^{0}_{\fant{a}3}$, $\Pi^{0}_{\fant{a}0}\leftrightarrow\Pi^{1}_{\fant{a}1}=-\Pi^{0}_{\fant{a}0}$, $\Pi^{3}_{\fant{a}3}\leftrightarrow\Pi^{2}_{\fant{a}2}=-\Pi^{3}_{\fant{a}3}$.
Likewise,  as an example of the prime symmetry for the QC part, one has the prime companions $E^1_{\fant{0}1}\leftrightarrow E^0_{\fant{0}0}$ which results from the exchange symmetry  $l\otimes l\leftrightarrow n\otimes n$, in the same way as $G^1_{\fant{a}1}$ and $G^0_{\fant{0}0}$ are the prime companion components of the Einstein  tensor for the usual $pp$-wave Ans\"atze in GR.

\section*{Acknowledgements}
The author would like to thank Professor Alikram N. Aliev for introducing him  to Newman-Penrose formalism. He is grateful to the anonymous referee
for the comments  that helped to improve the paper.

\end{document}